% EXTERNAL FILES

\input harvmac
\input amssym

% FONTS

% fraktur

\newfam\frakfam
\font\teneufm=eufm10
\font\seveneufm=eufm7
\font\fiveeufm=eufm5
\textfont\frakfam=\teneufm
\scriptfont\frakfam=\seveneufm
\scriptscriptfont\frakfam=\fiveeufm

% black board bold

\def\bb{
\font\tenmsb=msbm10
\font\sevenmsb=msbm7
\font\fivemsb=msbm5
\textfont1=\tenmsb
\scriptfont1=\sevenmsb
\scriptscriptfont1=\fivemsb
}

%\newfam\msbfam
%\font\tenmsb=msbm10
%\font\sevenmsb=msbm7
%\font\fivemsb=msbm5
%\textfont\msbfam=\tenmsb
%\scriptfont\msbfam=\sevenmsb
%\scriptscriptfont\msbfam=\fivemsb
%\def\bb{\fam\msbfam \tenmsb}

% double stroke math

\newfam\dsromfam
\font\tendsrom=dsrom10
\textfont\dsromfam=\tendsrom
\def\ds{\fam\dsromfam \tendsrom}

% bold math italics

\newfam\mbffam
\font\tenmbf=cmmib10
\font\sevenmbf=cmmib7
\font\fivembf=cmmib5
\textfont\mbffam=\tenmbf
\scriptfont\mbffam=\sevenmbf
\scriptscriptfont\mbffam=\fivembf

% bold math cal

\newfam\mbfcalfam
\font\tenmbfcal=cmbsy10
\font\sevenmbfcal=cmbsy7
\font\fivembfcal=cmbsy5
\textfont\mbfcalfam=\tenmbfcal
\scriptfont\mbfcalfam=\sevenmbfcal
\scriptscriptfont\mbfcalfam=\fivembfcal

% math script

\newfam\mscrfam
\font\tenmscr=rsfs10
\font\sevenmscr=rsfs7
\font\fivemscr=rsfs5
\textfont\mscrfam=\tenmscr
\scriptfont\mscrfam=\sevenmscr
\scriptscriptfont\mscrfam=\fivemscr

% MACROS

% bras, kets, ...

% tilde, hat, bar, ...

\def\tilde{\widetilde}

\def\bar{\overline}
\def\b{\bar}
\def\bsq#1{{{\b{#1}}^{\lower 2.5pt\hbox{$\scriptstyle 2$}}}}
\def\bexp#1#2{{{\b{#1}}^{\lower 2.5pt\hbox{$\scriptstyle #2$}}}}
\def\dotexp#1#2{{{#1}^{\lower 2.5pt\hbox{$\scriptstyle #2$}}}}

% basic math

\def\rt2{\sqrt{2}}
\def\half {{1 \over 2}}
\def\Re{\mathop{\rm Re}}

\def\Tr{\mathop{\rm Tr}}
\def\tr{\mathop{\rm tr}}

\def\underrel#1\over#2{\mathrel{\mathop{\kern\z@#1}\limits_{#2}}}

% bold greek characters

\font\tenbifull=cmmib10
\font\tenbimed=cmmib7
\font\tenbismall=cmmib5
\textfont9=\tenbifull \scriptfont9=\tenbimed
\scriptscriptfont9=\tenbismall

\mathchardef\bbGamma="7000
\mathchardef\bbDelta="7001
\mathchardef\bbPhi="7002
\mathchardef\bbAlpha="7003
\mathchardef\bbXi="7004
\mathchardef\bbPi="7005
\mathchardef\bbSigma="7006
\mathchardef\bbUpsilon="7007
\mathchardef\bbTheta="7008
\mathchardef\bbPsi="7009
\mathchardef\bbOmega="700A
\mathchardef\bbalpha="710B
\mathchardef\bbbeta="710C
\mathchardef\bbgamma="710D
\mathchardef\bbdelta="710E
\mathchardef\bbepsilon="710F
\mathchardef\bbzeta="7110
\mathchardef\bbeta="7111
\mathchardef\bbtheta="7112
\mathchardef\bbiota="7113
\mathchardef\bbkappa="7114
\mathchardef\bblambda="7115
\mathchardef\bbmu="7116
\mathchardef\bbnu="7117
\mathchardef\bbxi="7118
\mathchardef\bbpi="7119
\mathchardef\bbrho="711A
\mathchardef\bbsigma="711B
\mathchardef\bbtau="711C
\mathchardef\bbupsilon="711D
\mathchardef\bbphi="711E
\mathchardef\bbchi="711F
\mathchardef\bbpsi="7120
\mathchardef\bbomega="7121
\mathchardef\bbvarepsilon="7122
\mathchardef\bbvartheta="7123
\mathchardef\bbvarpi="7124
\mathchardef\bbvarrho="7125
\mathchardef\bbvarsigma="7126
\mathchardef\bbvarphi="7127

% dotted spinor indices

\def\nablaslash{\not{\hbox{\kern-2pt $\nabla$}}}

% bared indices

% bared spinors

% capital cal letters

\def\CA{{\cal A}}

\def\CH{{\cal H}}
\def\CI{{\cal I}}

\def\CL{{\cal L}}
\def\CM{{\cal M}}
\def\CN{{\cal N}}
\def\CO{{\cal O}}

\def\CR{{\cal R}}

\def\CW{{\cal W}}

% double stroke symbols: unit matrix, reals, complex, quaternions, integers, natural numbers

\def\1{{\ds 1}}
\def\R{\hbox{$\bb R$}}
\def\C{\hbox{$\bb C$}}

\def\Z{\hbox{$\bb Z$}}

% miscellaneous objects

%BIBLIOGRAPHY

%\KapustinGMA
\lref\KapustinGMA{
  A.~Kapustin,
  ``Bosonic Topological Insulators and Paramagnets: a view from cobordisms,''
[arXiv:1404.6659 [cond-mat.str-el]].
%%CITATION = arXiv:1404.6659%%
}

%\BanerjeeIZ
\lref\BanerjeeIZ{
  N.~Banerjee, J.~Bhattacharya, S.~Bhattacharyya, S.~Jain, S.~Minwalla and T.~Sharma,
  ``Constraints on Fluid Dynamics from Equilibrium Partition Functions,''
JHEP {\bf 1209}, 046 (2012).
[arXiv:1203.3544 [hep-th]].
%%CITATION = arXiv:1203.3544%%
}

 %\LandsteinerSJA
\lref\LandsteinerSJA{
  K.~Landsteiner,
  ``Anomaly related transport of Weyl fermions for Weyl semi-metals,''
Phys.\ Rev.\ B {\bf 89}, 075124 (2014).
[arXiv:1306.4932 [hep-th]].
%%CITATION = IFT-UAM-CSIC-13-074%%
}

%\LandsteinerKD
\lref\LandsteinerKD{
  K.~Landsteiner, E.~Megias and F.~Pena-Benitez,
  ``Anomalous Transport from Kubo Formulae,''
Lect.\ Notes Phys.\  {\bf 871}, 433 (2013).
[arXiv:1207.5808 [hep-th]].
%%CITATION = arXiv:1207.5808%%
}

%\HarveyBX
\lref\HarveyBX{
  J.~A.~Harvey, R.~Minasian and G.~W.~Moore,
  ``NonAbelian tensor multiplet anomalies,''
JHEP {\bf 9809}, 004 (1998).
[hep-th/9808060].
%%CITATION = hep-th/9808060%%
}

%\BardeenPM
\lref\BardeenPM{
  W.~A.~Bardeen and B.~Zumino,
  ``Consistent and Covariant Anomalies in Gauge and Gravitational Theories,''
Nucl.\ Phys.\ B {\bf 244}, 421 (1984).
%%CITATION = FERMILAB-PUB-84-038-T%%
}

%\JensenRGA
\lref\JensenRGA{
  K.~Jensen, R.~Loganayagam and A.~Yarom,
  ``Chern-Simons terms from thermal circles and anomalies,''
[arXiv:1311.2935 [hep-th]].
%%CITATION = YITP-SB-13-38%%
}

%\JensenKKA
\lref\JensenKKA{
  K.~Jensen, R.~Loganayagam and A.~Yarom,
  ``Anomaly inflow and thermal equilibrium,''
[arXiv:1310.7024 [hep-th]].
%%CITATION = YITP-SB-35%%
}

%\JensenKJ
\lref\JensenKJ{
  K.~Jensen, R.~Loganayagam and A.~Yarom,
  ``Thermodynamics, gravitational anomalies and cones,''
JHEP {\bf 1302}, 088 (2013).
[arXiv:1207.5824 [hep-th]].
%%CITATION = arXiv:1207.5824%%
}

%\GolkarKB
\lref\GolkarKB{
  S.~Golkar and D.~T.~Son,
  ``Non-Renormalization of the Chiral Vortical Effect Coefficient,''
[arXiv:1207.5806 [hep-th]].
%%CITATION = arXiv:1207.5806%%
}

%\DumitrescuHA
\lref\DumitrescuHA{
  T.~T.~Dumitrescu, G.~Festuccia and N.~Seiberg,
  ``Exploring Curved Superspace,''
JHEP {\bf 1208}, 141 (2012).
[arXiv:1205.1115 [hep-th]].
%%CITATION = arXiv:1205.1115%%
}

%\ClossetRU
\lref\ClossetRU{
  C.~Closset, T.~T.~Dumitrescu, G.~Festuccia and Z.~Komargodski,
  ``Supersymmetric Field Theories on Three-Manifolds,''
JHEP {\bf 1305}, 017 (2013).
[arXiv:1212.3388 [hep-th]].
%%CITATION = PUPT-2432%%
}

%\KuzenkoUYA
\lref\KuzenkoUYA{
  S.~M.~Kuzenko, U.~Lindstrom, M.~Rocek, I.~Sachs and G.~Tartaglino-Mazzucchelli,
  ``Three-dimensional N=2 supergravity theories: From superspace to components,''
Phys.\ Rev.\ D {\bf 89}, 085028 (2014).
[arXiv:1312.4267 [hep-th]].
%%CITATION = UUITP-20-13%%
}

%\ClossetVG
\lref\ClossetVG{
  C.~Closset, T.~T.~Dumitrescu, G.~Festuccia, Z.~Komargodski and N.~Seiberg,
  ``Contact Terms, Unitarity, and F-Maximization in Three-Dimensional Superconformal Theories,''
JHEP {\bf 1210}, 053 (2012).
[arXiv:1205.4142 [hep-th]].
%%CITATION = arXiv:1205.4142%%
}

%\ClossetVP
\lref\ClossetVP{
  C.~Closset, T.~T.~Dumitrescu, G.~Festuccia, Z.~Komargodski and N.~Seiberg,
  ``Comments on Chern-Simons Contact Terms in Three Dimensions,''
JHEP {\bf 1209}, 091 (2012).
[arXiv:1206.5218 [hep-th]].
%%CITATION = arXiv:1206.5218%%
}

%\DHA
\lref\AharonyDHA{
  O.~Aharony, S.~S.~Razamat, N.~Seiberg and B.~Willett,
  ``3d dualities from 4d dualities,''
JHEP {\bf 1307}, 149 (2013).
[arXiv:1305.3924 [hep-th]].
%%CITATION = WIS-04-13-APR-DPPA%%
}

%\KimQF
\lref\KimQF{
  H.~-C.~Kim, J.~Kim and S.~Kim,
  ``Instantons on the 5-sphere and M5-branes,''
[arXiv:1211.0144 [hep-th]].
%%CITATION = arXiv:1211.0144%%
}

%\KlareGN
\lref\KlareGN{
  C.~Klare, A.~Tomasiello and A.~Zaffaroni,
  ``Supersymmetry on Curved Spaces and Holography,''
JHEP {\bf 1208}, 061 (2012).
[arXiv:1205.1062 [hep-th]].
%%CITATION = arXiv:1205.1062%%
}

%\DolanQI
\lref\DolanQI{
  F.~A.~Dolan and H.~Osborn,
  ``Applications of the Superconformal Index for Protected Operators and q-Hypergeometric Identities to N=1 Dual Theories,''
Nucl.\ Phys.\ B {\bf 818}, 137 (2009).
[arXiv:0801.4947 [hep-th]].
%%CITATION = arXiv:0801.4947%%
}

%\BonettiFN
\lref\BonettiFN{
  F.~Bonetti, T.~W.~Grimm and S.~Hohenegger,
  ``A Kaluza-Klein inspired action for chiral p-forms and their anomalies,''
Phys.\ Lett.\ B {\bf 720}, 424 (2013).
[arXiv:1206.1600 [hep-th]].
%%CITATION = arXiv:1206.1600%%
}

%\DolanRP
\lref\DolanRP{
  F.~A.~H.~Dolan, V.~P.~Spiridonov and G.~S.~Vartanov,
  ``From 4d superconformal indices to 3d partition functions,''
Phys.\ Lett.\ B {\bf 704}, 234 (2011).
[arXiv:1104.1787 [hep-th]].
%%CITATION = arXiv:1104.1787%%
}

%\ImamuraUW
\lref\ImamuraUW{
  Y.~Imamura,
  ``Relation between the 4d superconformal index and the $S^3$ partition function,''
JHEP {\bf 1109}, 133 (2011).
[arXiv:1104.4482 [hep-th]].
%%CITATION = arXiv:1104.4482%%
}

%\FestucciaWS
\lref\FestucciaWS{
  G.~Festuccia and N.~Seiberg,
  ``Rigid Supersymmetric Theories in Curved Superspace,''
JHEP {\bf 1106}, 114 (2011).
[arXiv:1105.0689 [hep-th]].
%%CITATION = arXiv:1105.0689%%
}

%\CardyIE
\lref\CardyIE{
  J.~L.~Cardy,
  ``Operator Content of Two-Dimensional Conformally Invariant Theories,''
Nucl.\ Phys.\ B {\bf 270}, 186 (1986).
}

%\KinneyEJ
\lref\KinneyEJ{
  J.~Kinney, J.~M.~Maldacena, S.~Minwalla and S.~Raju,
  ``An Index for 4 dimensional super conformal theories,''
Commun.\ Math.\ Phys.\  {\bf 275}, 209 (2007).
[hep-th/0510251].
%%CITATION = hep-th/0510251%%
}

%\RomelsbergerEG
\lref\RomelsbergerEG{
  C.~Romelsberger,
  ``Counting chiral primaries in $\CN = 1$, $d=4$ superconformal field theories,''
Nucl.\ Phys.\ B {\bf 747}, 329 (2006).
[hep-th/0510060].
%%CITATION = hep-th/0510060%%
}

%\MartelliFU
\lref\MartelliFU{
  D.~Martelli, A.~Passias and J.~Sparks,
  ``The gravity dual of supersymmetric gauge theories on a squashed three-sphere,''
Nucl.\ Phys.\ B {\bf 864}, 840 (2012).
[arXiv:1110.6400 [hep-th]].
%%CITATION = arXiv:1110.6400%%
}

%\ImamuraUW
\lref\ImamuraUW{
  Y.~Imamura,
  ``Relation between the 4d superconformal index and the $S^3$ partition function,''
JHEP {\bf 1109}, 133 (2011).
[arXiv:1104.4482 [hep-th]].
%%CITATION = arXiv:1104.4482%%
}

%\SpiridonovWW
\lref\SpiridonovWW{
  V.~P.~Spiridonov and G.~S.~Vartanov,
  ``Elliptic hypergeometric integrals and 't Hooft anomaly matching conditions,''
JHEP {\bf 1206}, 016 (2012).
[arXiv:1203.5677 [hep-th]].
%%CITATION = DESY-12-048%%
}

%\GubserNZ
\lref\GubserNZ{
  S.~S.~Gubser, I.~R.~Klebanov and A.~A.~Tseytlin,
  ``Coupling constant dependence in the thermodynamics of N=4 supersymmetric Yang-Mills theory,''
Nucl.\ Phys.\ B {\bf 534}, 202 (1998).
[hep-th/9805156].
%%CITATION = hep-th/9805156%%
}

%\ClossetVRA
\lref\ClossetVRA{
  C.~Closset, T.~T.~Dumitrescu, G.~Festuccia and Z.~Komargodski,
  ``The Geometry of Supersymmetric Partition Functions,''
JHEP {\bf 1401}, 124 (2014).
[arXiv:1309.5876 [hep-th]].
%%CITATION = WIS-09-13-SEP-DPPA%%
}

%\WittenDF
\lref\WittenDF{
  E.~Witten,
  ``Constraints on Supersymmetry Breaking,''
Nucl.\ Phys.\ B {\bf 202}, 253 (1982).
%%CITATION = PRINT-82-0163 (PRINCETON)%%
}

%\SonTF
\lref\SonTF{
  D.~T.~Son and P.~Surowka,
  ``Hydrodynamics with Triangle Anomalies,''
Phys.\ Rev.\ Lett.\  {\bf 103}, 191601 (2009).
[arXiv:0906.5044 [hep-th]].
%%CITATION = INT-PUB-09-030%%
}

%\LandsteinerCP
\lref\LandsteinerCP{
  K.~Landsteiner, E.~Megias and F.~Pena-Benitez,
  ``Gravitational Anomaly and Transport,''
Phys.\ Rev.\ Lett.\  {\bf 107}, 021601 (2011).
[arXiv:1103.5006 [hep-ph]].
%%CITATION = arXiv:1103.5006%%
}

%\AnselmiAM
\lref\AnselmiAM{
  D.~Anselmi, D.~Z.~Freedman, M.~T.~Grisaru and A.~A.~Johansen,
  ``Nonperturbative formulas for central functions of supersymmetric gauge theories,''
Nucl.\ Phys.\ B {\bf 526}, 543 (1998).
[hep-th/9708042].
%%CITATION = hep-th/9708042%%
}

%\LoganayagamPZ
\lref\LoganayagamPZ{
  R.~Loganayagam and P.~Surowka,
  ``Anomaly/Transport in an Ideal Weyl gas,''
JHEP {\bf 1204}, 097 (2012).
[arXiv:1201.2812 [hep-th]].
%%CITATION = arXiv:1201.2812%%
}

%\LandsteinerIQ
\lref\LandsteinerIQ{
  K.~Landsteiner, E.~Megias, L.~Melgar and F.~Pena-Benitez,
  ``Holographic Gravitational Anomaly and Chiral Vortical Effect,''
JHEP {\bf 1109}, 121 (2011).
[arXiv:1107.0368 [hep-th]].
%%CITATION = arXiv:1107.0368%%
}

%\ColemanZI
\lref\ColemanZI{
  S.~R.~Coleman and B.~R.~Hill,
  ``No More Corrections to the Topological Mass Term in QED in Three-Dimensions,''
Phys.\ Lett.\ B {\bf 159}, 184 (1985).
%%CITATION = HUTP-85-A047%%
}

%\KutasovSV
\lref\KutasovSV{
  D.~Kutasov and N.~Seiberg,
  ``Number of degrees of freedom, density of states and tachyons in string theory and CFT,''
Nucl.\ Phys.\ B {\bf 358}, 600 (1991).
%%CITATION = PUPT-1221%%
}

%\BuicanQLA
\lref\BuicanQLA{
  M.~Buican, T.~Nishinaka and C.~Papageorgakis,
  ``Constraints on Chiral Operators in N=2 SCFTs,''
[arXiv:1407.2835 [hep-th]].
%%CITATION = arXiv:1407.2835%%
}

%\SpiridonovDE
\lref\SpiridonovDE{
  V.~P.~Spiridonov,
  ``Modified elliptic gamma functions and 6d superconformal indices,''
Lett.\ Math.\ Phys.\  {\bf 104}, 397 (2014).
[arXiv:1211.2703 [hep-th]].
%%CITATION = arXiv:1211.2703%%
}

%\AlvarezGaumeIG
\lref\AlvarezGaumeIG{
  L.~Alvarez-Gaume and E.~Witten,
  ``Gravitational Anomalies,''
Nucl.\ Phys.\ B {\bf 234}, 269 (1984).
%%CITATION = HUTP-83/A039%%
}

%\ImamuraXG
\lref\ImamuraXG{
  Y.~Imamura,
  ``Supersymmetric theories on squashed five-sphere,''
PTEP {\bf 2013}, 013B04 (2013).
[arXiv:1209.0561 [hep-th]].
%%CITATION = arXiv:1209.0561%%
}

%\ImamuraXNA
\lref\ImamuraXNA{
  Y.~Imamura,
  ``Perturbative partition function for a squashed S5,''
PTEP {\bf 2013}, no. 7, 073B01 (2013).
}

%\CamanhoAPA
\lref\CamanhoAPA{
  X.~O.~Camanho, J.~D.~Edelstein, J.~Maldacena and A.~Zhiboedov,
  ``Causality Constraints on Corrections to the Graviton Three-Point Coupling,''
[arXiv:1407.5597 [hep-th]].
%%CITATION = arXiv:1407.5597%%
}

%\LockhartVP
\lref\LockhartVP{
  G.~Lockhart and C.~Vafa,
  ``Superconformal Partition Functions and Non-perturbative Topological Strings,''
[arXiv:1210.5909 [hep-th]].
%%CITATION = arXiv:1210.5909%%
}

%\AldayBTA
\lref\AldayBTA{
  L.~F.~Alday, M.~Fluder, C.~M.~Gregory, P.~Richmond and J.~Sparks,
  ``Supersymmetric gauge theories on squashed five-spheres and their gravity duals,''
[arXiv:1405.7194 [hep-th]].
%%CITATION = arXiv:1405.7194%%
}

%\CassaniZWA
\lref\CassaniZWA{
  D.~Cassani and D.~Martelli,
  ``The gravity dual of supersymmetric gauge theories on a squashed $S^1 \times S^3$,''
[arXiv:1402.2278 [hep-th]].
%%CITATION = arXiv:1402.2278%%
}

%\AlvarezGaumeDR
\lref\AlvarezGaumeDR{
  L.~Alvarez-Gaume and P.~H.~Ginsparg,
  ``The Structure of Gauge and Gravitational Anomalies,''
Annals Phys.\  {\bf 161}, 423 (1985), [Erratum-ibid.\  {\bf 171}, 233 (1986)].
%%CITATION = HUTP-84/A016%%
}

%\BonettiELA
\lref\BonettiELA{
  F.~Bonetti, T.~W.~Grimm and S.~Hohenegger,
  ``One-loop Chern-Simons terms in five dimensions,''
JHEP {\bf 1307}, 043 (2013).
[arXiv:1302.2918 [hep-th]].
%%CITATION = arXiv:1302.2918%%
}

%\AldayRXA
\lref\AldayRXA{
  L.~F.~Alday, M.~Fluder, P.~Richmond and J.~Sparks,
  ``The gravity dual of supersymmetric gauge theories on a squashed five-sphere,''
[arXiv:1404.1925 [hep-th]].
%%CITATION = arXiv:1404.1925%%
}

%\BhattacharyaZY
\lref\BhattacharyaZY{
  J.~Bhattacharya, S.~Bhattacharyya, S.~Minwalla and S.~Raju,
  ``Indices for Superconformal Field Theories in 3,5 and 6 Dimensions,''
JHEP {\bf 0802}, 064 (2008).
[arXiv:0801.1435 [hep-th]].
%%CITATION = arXiv:0801.1435%%
}

%\NiemiRQ
\lref\NiemiRQ{
  A.~J.~Niemi and G.~W.~Semenoff,
  ``Axial Anomaly Induced Fermion Fractionization and Effective Gauge Theory Actions in Odd Dimensional Space-Times,''
Phys.\ Rev.\ Lett.\  {\bf 51}, 2077 (1983).
%%CITATION = Print-83-0988 (IAS,PRINCETON)%%
}

%\RedlichKN
\lref\RedlichKN{
  A.~N.~Redlich,
  ``Gauge Noninvariance and Parity Violation of Three-Dimensional Fermions,''
Phys.\ Rev.\ Lett.\  {\bf 52}, 18 (1984).
%%CITATION = MIT-CTP-1107%%
}

%\KimAVA
\lref\KimAVA{
  H.~-C.~Kim and S.~Kim,
  ``M5-branes from gauge theories on the 5-sphere,''
JHEP {\bf 1305}, 144 (2013).
[arXiv:1206.6339 [hep-th]].
%%CITATION = SNUTP12-002%%
}

%\RedlichDV
\lref\RedlichDV{
  A.~N.~Redlich,
  ``Parity Violation and Gauge Noninvariance of the Effective Gauge Field Action in Three-Dimensions,''
Phys.\ Rev.\ D {\bf 29}, 2366 (1984).
%%CITATION = MIT-CTP-1128%%
}

%\KimAVA
\lref\KimAVA{
  H.~-C.~Kim and S.~Kim,
  ``M5-branes from gauge theories on the 5-sphere,''
JHEP {\bf 1305}, 144 (2013).
[arXiv:1206.6339 [hep-th]].
%%CITATION = SNUTP12-002%%
}

%\DumitrescuIU
\lref\DumitrescuIU{
  T.~T.~Dumitrescu and N.~Seiberg,
  ``Supercurrents and Brane Currents in Diverse Dimensions,''
JHEP {\bf 1107}, 095 (2011).
[arXiv:1106.0031 [hep-th]].
%%CITATION = arXiv:1106.0031%%
}

%\KomargodskiRB
\lref\KomargodskiRB{
  Z.~Komargodski and N.~Seiberg,
  ``Comments on Supercurrent Multiplets, Supersymmetric Field Theories and Supergravity,''
JHEP {\bf 1007}, 017 (2010).
[arXiv:1002.2228 [hep-th]].
%%CITATION = arXiv:1002.2228%%
}

%\AldayLBA
\lref\AldayLBA{
  L.~F.~Alday, D.~Martelli, P.~Richmond and J.~Sparks,
  ``Localization on Three-Manifolds,''
JHEP {\bf 1310}, 095 (2013).
[arXiv:1307.6848 [hep-th]].
%%CITATION = arXiv:1307.6848%%
}

%\KimNVA
\lref\KimNVA{
  H.~-C.~Kim, S.~Kim, S.~-S.~Kim and K.~Lee,
  ``The general M5-brane superconformal index,''
[arXiv:1307.7660].
%%CITATION = KIAS-P13039%%
}

%\IntriligatorEQ
\lref\IntriligatorEQ{
  K.~A.~Intriligator,
  ``Anomaly matching and a Hopf-Wess-Zumino term in 6d, N=(2,0) field theories,''
Nucl.\ Phys.\ B {\bf 581}, 257 (2000).
[hep-th/0001205].
%%CITATION = hep-th/0001205%%
}

%\CardyIE
\lref\CardyIE{
  J.~L.~Cardy,
  ``Operator Content of Two-Dimensional Conformally Invariant Theories,''
Nucl.\ Phys.\ B {\bf 270}, 186 (1986).
}

%\SohniusTP
\lref\SohniusTP{
  M.~F.~Sohnius and P.~C.~West,
  ``An Alternative Minimal Off-Shell Version of N=1 Supergravity,''
Phys.\ Lett.\ B {\bf 105}, 353 (1981).
%%CITATION = ICTP 80-81/37%%
}

%\ArdehaliXYA
\lref\ArdehaliXYA{
  A.~A.~Ardehali, J.~T.~Liu and P.~Szepietowski,
  ``$1/N^2$ corrections to the holographic Weyl anomaly,''
JHEP {\bf 1401}, 002 (2014).
[arXiv:1310.2611 [hep-th]].
%%CITATION = MCTP-13-28%%
}

%\MaxfieldAW
\lref\MaxfieldAW{
  T.~Maxfield and S.~Sethi,
  ``The Conformal Anomaly of M5-Branes,''
JHEP {\bf 1206}, 075 (2012).
[arXiv:1204.2002 [hep-th]].
%%CITATION = arXiv:1204.2002%%
}

%\ArdehaliXYA
\lref\ArdehaliToAppear{
  A.~A.~Ardehali, J.~T.~Liu and P.~Szepietowski, ``$c - a$ from the $\CN = 1$ superconformal index.'' To Appear.
%%CITATION = MCTP-13-28%%
}

%\ArdehaliZBA
\lref\ArdehaliZBA{
  A.~A.~Ardehali, J.~T.~Liu and P.~Szepietowski,
  ``c-a from the N=1 superconformal index,''
[arXiv:1407.6024 [hep-th]].
%%CITATION = MCTP-14-23%%
}

%\ClossetSXA
\lref\ClossetSXA{
  C.~Closset and I.~Shamir,
  ``The $\CN=1$ Chiral Multiplet on $T^2\times S^2$ and Supersymmetric Localization,''
JHEP {\bf 1403}, 040 (2014).
[arXiv:1311.2430 [hep-th]].
%%CITATION = WIS-12-13-OCT-DPPA%%
}

%\AldayRXA
\lref\AldayRXA{
  L.~F.~Alday, M.~Fluder, P.~Richmond and J.~Sparks,
  ``The gravity dual of supersymmetric gauge theories on a squashed five-sphere,''
[arXiv:1404.1925 [hep-th]].
%%CITATION = arXiv:1404.1925%%
}

%\AsselPAA
\lref\AsselPAA{
  B.~Assel, D.~Cassani and D.~Martelli,
  ``Localization on Hopf surfaces,''
[arXiv:1405.5144 [hep-th]].
%%CITATION = KCL-MTH-14-09%%
}

%\GerchkovitzZRA
\lref\GerchkovitzZRA{
  E.~Gerchkovitz,
  ``Constraints on the R-charges of Free Bound States from the R\"omelsberger Index,''
[arXiv:1311.0487 [hep-th]].
%%CITATION = arXiv:1311.0487%%
}

%\KlebanovGS
\lref\KlebanovGS{
  I.~R.~Klebanov, S.~S.~Pufu and B.~R.~Safdi,
  ``F-Theorem without Supersymmetry,''
JHEP {\bf 1110}, 038 (2011).
[arXiv:1105.4598 [hep-th]].
%%CITATION = arXiv:1105.4598%%
}

%\ClossetUDA
\lref\ClossetUDA{
  C.~Closset, T.~T.~Dumitrescu, G.~Festuccia and Z.~Komargodski,
  ``From Rigid Supersymmetry to Twisted Holomorphic Theories,''
[arXiv:1407.2598 [hep-th]].
%%CITATION = WIS-05-14-JUN-DPPA%%
}

%\AsselInProgress
\lref\AsselInProgress{
  B.~Assel, D.~Cassani, L.~Di~Pietro, Z.~Komargodski, J.~Lorenzen, D.~Martelli,
  in progress.
%%CITATION = WIS-05-14-JUN-DPPA%%
}

%BODY

%\draftmode

\rightline{WIS/06/14-JUL-DPPA}
\vskip-20pt
\Title{
} {\vbox{\centerline{Cardy Formulae for SUSY Theories in $d=4$ and $d=6$}
 }}

\vskip-15pt
\centerline{Lorenzo Di Pietro and Zohar Komargodski}
\vskip15pt
  \centerline{\it Weizmann Institute of Science, Rehovot
76100, Israel}

\vskip35pt

\noindent

We consider supersymmetric theories on a space with compact space-like slices. One can count BPS representations weighted by $(-1)^F$, or, equivalently, study supersymmetric partition functions by compactifying the time direction. A special case of this general construction corresponds to the counting of short representations of the  superconformal group.
We show that in four-dimensional $\CN=1$ theories  the ``high temperature'' asymptotics of such counting problems is  fixed by the anomalies of the theory. Notably, the combination $a-c$ of the trace anomalies plays a crucial role. We also propose similar formulae for six-dimensional $(1,0)$ theories.

\Date{July 2014}

\newsec{Introduction}

Since the work of Witten~\WittenDF\ it has been clear that in some situations non-perturbative computations in supersymmetric theories can be performed at weak (or even zero) coupling. Suppose we are given a supercharge $Q$ with $\{Q,Q^\dagger\}=\Delta$ where $\Delta$ is some conserved  charge.  Let the Hilbert space be $\CH$, then we may consider the following index:
\eqn\genindex{\CI[\mu_i]={\Tr}_{\CH}\left[(-1)^F\prod_i z_i^{ q_i}  \right]~.}
$F$ is a $\Z_2$-valued fermion number operator such that $[F,Q]=-Q$. $q_i$ stand for conserved charges that commute with $Q$, $[q_i,Q]=0$ (therefore, they also commute with $\Delta$) and $z_i $ are the fugacities associated with the $q_i$. 

From the above, it follows that only states with $\Delta=0$ contribute to $\CI$. The next key observation is that representations of the algebra $\{Q,Q^\dagger\}=0$ are short compared to the case that $\Delta\neq 0$. Finally, two short representations can combine to a long representation only if they have different fermion numbers. 

Therefore, the trace~\genindex\ is independent of continuous coupling constants and it can be often computed at zero coupling.\foot{One has to make sure not to introduce coupling constants which modify the Hilbert space too radically. See for example~\GerchkovitzZRA\ for a discussion that also pertains to some of the cases of interest below.} An obstacle to the idea of  carrying out such computations in many interesting supersymmetric quantum field theories is that in supersymmetric theories one often has a continuous moduli space of vacua. Then, the states in the kernel of $\Delta$ form a continuum and it is not clear how to define~\genindex. 

A possible way to proceed is to study the theory on $\CM_{d-1}\times \R$ with $\CM_{d-1}$ some compact $d-1$-dimensional manifold.  Since on curved spaces one often finds that the scalar fields are coupled to curvature, one may hope that the continuous moduli space is lifted. If so, the index~\genindex\ can be computed~\RomelsbergerEG\ and it is an interesting object to study. As an example, we could take dual theories and compactify them on the same space $\CM_{d-1}\times \R$. Then, a nontrivial check  of the duality would be that the indices for the dual theories agree.

Not every choice of $\CM_{d-1}$ is consistent with preserving some supersymmetry. We will recall the conditions that $\CM_{d-1}$ needs to satisfy in order for it to be consistent with unbroken supersymmetry.

A particularly interesting choice to make is $\CM_{d-1}=S^{d-1}$. Further, we can take the field theory to be superconformal. Then, since $S^{d-1}\times \R$  is conformally flat, the index~\genindex\ in this case can be related via radial quantization to  counting local operators in $\R^d$ that sit in short representations of the  superconformal group. 

One of the natural fugacities that one can introduce in supersymmetric field theories on~$S^{d-1}\times \R$ geometrically corresponds to compactifying $\R\to S^1$.  In other words, we can consider spaces of the form $\CM_d = \CM_{d-1} \times S^1$. If we denote the generator of translations along the $S^1$ by $H$, then the partition function over $\CM_d$ can be interpreted as
 \eqn\statsum{Z_{\CM_d}= {\Tr}_{\CH(\CM_{d-1})} \left[(-1)^Fe^{-\beta\left(H - \sum_i \mu_i q_i \right)}\right]~.}
Here we have assumed that the fermions have periodic boundary conditions. The length of the $S^1$ is $\beta\equiv 2\pi r_1\equiv T^{-1}$. We have also allowed for various chemical potentials $\mu_i$ that couple to conserved charges $q_i$ which commute with the SUSY generators on $\CM_{d-1}\times S^1$.

Most of our discussion will be in the context of $\CN=1$ $d=4$ theories. The case of $(1,0)$ theories in $d=6$ is  analogous. For technical reasons, everywhere below we assume the existence of an $R$-symmetry.

First of all, some preliminaries:  An interesting family of  spaces $\CM_3\times \R$ which admit unbroken SUSY generators which do not depend on time (i.e. do not depend on the coordinate of $\R$) is obtained by taking $\CM_3$ to be a Seifert manifold~\refs{\KlareGN, \DumitrescuHA,\ClossetRU}. A Seifert manifold is simply an $S^1$ fibration over a Riemann surface. (Some simple examples in this class are therefore $S^2\times S^1$, $S^3$, and Lens spaces.) Such spaces preserve at least two supersymmetry generators $\delta_\zeta$ and $\delta_{\tilde\zeta}$ of opposite $R$-charge. Since the supersymmetry generators are time-independent, there is no obstruction to compactifying $\R\to S^1$ and we can thus consider $\CM_4=\CM_3\times S^1$ with $\CM_3$ any Seifert manifold. 

The total four-dimensional space $\CM_4=\CM_3\times S^1$ is then guaranteed to be a complex manifold,\foot{In most generality (which we will not need in this paper), for $R$-symmetric $\CN=1$ theories in four dimensions,  a four-manifold $\CM_4$ can preserve at least one supersymmetry generator if and only if it is a complex manifold~\refs{\KlareGN, \DumitrescuHA}.} and there is a holomorphic Killing vector that points in a direction which is a linear combination of the Seifert circle and the $S^1$ in $\CM_3\times S^1$. Let us call this holomorphic Killing vector $K$. The SUSY algebra is then 
\eqn\superalgebraintro{\{\delta_\zeta,\delta_{\tilde \zeta}\}=2i\delta_K~,\qquad \delta_\zeta^2=\delta_{\tilde \zeta}^2=0~.}

We will show that when the $S^1$ becomes small, i.e. $\beta\to 0$,  the leading contribution to the supersymmetric partition function~\statsum\ with $\mu_i=0$ is universal and takes the form
\eqn\CardySUSY{\beta \to 0: \quad \log Z_{\CM_4} \sim -{\pi^2  \Tr(R)  L_{\CM_3} \over \beta}~,}
where $L_{\CM_3}$ is a length scale associated to the manifold $\CM_3$. We will provide an explicit formula for $L_{\CM_3}$ that can be used to evaluate it on any Seifert manifold. $\Tr(R)$ is the mixed  anomaly of the $R$-symmetry of the theory. Note that a volume term (i.e. a term that goes like $\beta^{-3}$) is absent from~\CardySUSY. This is the usual statement that supersymmetric theories do not generate a cosmological constant. 

If the theory is superconformal and $\CM_3=S^3$, then, by the state-operator correspondence, $\CH(S^3)$ is  isomorphic to the space of local operators of the SCFT in $\R^4$. The generator $H$ which commutes with the SUSY algebra is a combination of the conformal dimension $\Delta$ and the $R$-charge $R$, $H=\Delta+\half R$. Recall that the $R$-symmetry that sits in the superconformal group satisfies~\AnselmiAM\ $\Tr(R)=16(a-c)$. Then, from~\CardySUSY\ we find that
\eqn\CardySCFT{\beta \to 0:\ \ \ \sum_{operators}(-1)^F e^{-\beta\left(\Delta+1/2 R\right)}\sim e^{ -{16 \pi^2    \over 3\beta}(a-c)}~,}
where we have taken the radius of the $S^3$ to be one. Only operators that sit in short representations of the superconformal group contribute to the left hand side of~\CardySCFT, see~\KinneyEJ.
Therefore,~\CardySCFT\ encodes  a universal property of the spectral density of ``heavy'' BPS operators in  $\CN=1$ SCFTs in four dimensions. As in the more general case~\CardySUSY, equation~\CardySCFT\ can be refined to include a squashing parameter of $S^3$.  Note that~\CardySCFT\ implies that $a-c$ can be computed just from the spectrum of BPS operators. (Conventionally, the $a$- and $c$- anomalies are extracted from three-point functions of the energy-momentum supermultiplet.) A statement that seems to be related was recently made in the context of holography~\ArdehaliXYA, see also the more explicit suggestion of~\ArdehaliZBA.

Equation~\CardySCFT\ is reminiscent of Cardy's universal formula~\CardyIE\ in two dimensions 
\eqn\Cardy{\beta \to 0:\quad \sum_{operators}e^{-\beta\Delta} \sim { e^{\pi^2 c \over 3  \beta}}~.  }
where  $c$ is the Virasoro central charge and the spatial circle is again taken to have radius one. Equation~\Cardy\ is intimately related to the modular group in two dimensions. It would be nice to understand if the analogy to~\CardySCFT\ thus suggests the existence of similar structures in $d>2$, at least for this class of supersymmetric observables.

Six-dimensional $(1,0)$ theories can be studied on $S^5\times S^1$ while preserving supersymmetry. While we are able to prove~\CardySCFT\ in quite some generality, some of the necessary  preliminary work on supersymmetry in six (and five) dimensions has not been carried out yet. Nevertheless, we  use more heuristic methods that lead to the following universal proposal for the $\beta\rightarrow 0$ limit of the partition function on $S^5\times S^1$ (with $\omega_{1,2,3}$ being squashing parameter~\ImamuraXNA\ (and $r_5$ being the radius of the $S^5$):
\eqn\limitindsixintro{\eqalign{\beta \to 0: & \quad  \log Z_{S^5 \times S^1} \sim - {\pi^2 \over \omega_1 \omega_2 \omega_3 } \left( {\kappa_1 \pi^2  \over 45 }{r_5^3 \over \beta^3} \right. \cr & \left.  + {(\omega_1^2 + \omega_2^2 + \omega_3^2) \kappa_2 + 3 (\omega_1 \omega_2 + \omega_2 \omega_3 + \omega_3 \omega_1)\kappa_3 \over 36 }{r_5 \over \beta}  \right)~,} }
where $\kappa_{1,2,3}$ are theory-dependent coefficients that are related to the coefficients in the eight-form anomaly polynomial ($R$ is the Riemann curvature two-form, $F$ is the field strength of the $SU(2)_R$ symmetry)
\eqn\anopolsixintro{I_8 = {1 \over (4 \pi)^3} \left[ {A \over 720} \tr R^4 + {B \over 576} (\tr R^2)^2 - {C\over 12} \tr R^2 \Tr F^2 + {D \over 3} \Tr F^4 \right]~,}
by
\eqn\conjectureintro{\kappa_1 = -{A + 5 B \over 6}\,,\quad \kappa_2  = {A + 2 B  \over 6}-{3\over 2} C\,,\quad \kappa_3 = -C ~.}
Our normalization is such that $\kappa_1 = \kappa_2 = \kappa_3 = 1$ for a free vector multiplet. 

We will now explain briefly how we  obtain~\CardySUSY\ and~\limitindsixintro.
It is helpful to start from the usual thermal partition function of QFT (not necessarily supersymmetric). We thus consider an arbitrary QFT on the space $\CM_3\times S^1$ with the fermions  assigned anti-periodic boundary conditions along the circle. This partition function captures the equilibrium properties of the quantum field theory at finite temperature $T = \beta^{-1} \equiv (2\pi r_1)^{-1}$. 
Further, let us assume the theory has a conserved $U(1)$ symmetry, with $q$ being the corresponding charge. It is useful to introduce a background  metric $g_{\mu\nu}$ that couples to the energy-momentum tensor and a background gauge field $A_\mu$ that couples to the conserved current. In order to obtain correlation functions at zero Matsubara frequency, one can reduce over the $S^1$ and find a local three-dimensional functional, $\CW$, of the background metric and gauge field. Derivatives of $\CW$ with respect to the background fields generate equilibrium correlation functions of the energy-momentum tensor and the conserved current.
 The expansion in derivatives of $\CW$ corresponds to the expansion in the radius of $S^1$ compared to the radius of $\CM_3$. (If $\CM_3=\R^3$ then the expansion in derivatives is just the usual expansion in the 3d momentum relative to the plasma.)

The functional $\CW$ of the background fields contains the following Chern-Simons (CS) term 
\eqn\mixedCS{\CW \supset {1 \over r_1}{i k \over 4\pi}\int_{\CM_3} A \wedge d a~,}
where $a$ is the KK photon (i.e. the mixed components of the metric tensor) and $A$ is the three-dimensional gauge field.  The coefficient $k$ has been calculated in several examples (both in free theories \LandsteinerCP~and at strong coupling using holography \LandsteinerIQ) and found to be proportional to the coefficient of the mixed $U(1)$-gravitational anomaly
\eqn\gravanoCS{k = -{1 \over 12} \Tr(q)~.}
This relation was conjectured to hold in general in \LandsteinerIQ~and generalized to higher (even) dimensions in \LoganayagamPZ.
Evidence for the robustness of \gravanoCS~in the context of perturbation theory was given in \GolkarKB. A non-perturbative explanation was presented in \JensenKJ~(and further studied and generalized in \JensenKKA,\JensenRGA) by assuming some regularity properties of the partition function on background geometries with conical singularities. In the main text we will give a very simple  non-perturbative derivation of \gravanoCS\ in a large class of theories.  

In the context of supersymmetric theories, the local term~\mixedCS\ needs to be  supersymmetrized (one needs to remember that~\gravanoCS\ is derived for anti-periodic boundary conditions, so this introduces a small complication that we will address). When the $U(1)$ symmetry is the $R$-symmetry, one finds that it is connected to the Einstein-Hilbert term on $\CM_3$ as well as to some other terms that come from the appropriate off-shell supergravity multiplet in three-dimensions. This local term contributes nontrivially to the partition function on curved manifolds. Since this is the leading term that appears in the derivative expansion on $\CM_3$, we obtain~\CardySUSY.

For six-dimensional theories there are more Chern-Simons terms that appear, and they again need to be appropriately supersymmetrized. The suitable off-shell supergravity formalism is not yet available in sufficient detail, so we do not have a comparably complete picture. However, by just considering the possible terms that can appear upon supersymmetrization and appealing to dimensional analysis, one can obtain enough information to propose the result~\limitindsixintro.

The rest of this note is organized as follows. In section 2 we discuss the relevant aspects of thermal field theory and hydrodynamics in the presence of anomalies. We consider in detail the example of a free Weyl fermion and give a non-perturbative derivation of~\gravanoCS. In section 3 we discuss supersymmetric theories and prove~\CardySUSY. In section 4 we present  some consistency checks and applications of~\CardySUSY. In section~5 we outline the arguments leading to the proposal~\limitindsixintro.

\newsec{The Thermal Derivative Expansion and Anomalies}

\subsec{Preliminaries}

It is generally expected that if a physical system is brought into contact with a thermal bath, then the system develops a gap. While many counter-examples to this are known, in the following we will assume that this scenario indeed holds (in particular, we assume that the global symmetries we discuss below are not spontaneously broken). More precisely, for the following discussion to hold true, we only need to assume that at some large enough temperature there are no gapless modes. 

To study the equilibrium correlation functions of the theory, we place the theory on the Euclidean space $\R^3\times S^1$ with $S^1$ of length $\beta\equiv {1\over T}\equiv 2\pi r_1$. The boundary conditions for the fermions are taken to be anti-periodic. In this way, the path integral over $\R^3\times S^1$  represents the usual thermal partition function 
\eqn\path{Z_{ \R^3 \times S^1}=\Tr (e^{-\beta H})~.}
If we took the fermions to be periodic, then the partition function would correspond to an insertion of $(-1)^F$ in the right hand side (with $F$ being the $\Z_2$-valued fermion number).

One may also be interested in studying equilibrium correlation functions. 
This can be done by coupling the theory on $\R^3\times S^1$ to background fields. In particular, if we are interested in correlation functions of some conserved current, $j^\mu$, we could couple it to a background gauge field $A_\mu$ in the usual way $\delta S\sim \int_{\R^3\times S^1} d^4x A_\mu j^\mu+\CO(A^2)$. Similarly, we can study correlation functions of the energy-momentum tensor by coupling the theory to a background metric field.\foot{We therefore implicitly assume that the relevant flavor charges  as well as the Poincar\'e charges are derived from  well defined (i.e. gauge invariant) local conserved currents and energy-momentum tensor, respectively. This means that there exists a sensible notion of charge density and energy density. In some peculiar examples this is not the case, for example, in the theory of the free gravitino $\Psi_{\mu\alpha}$, there is no conserved axial current  invariant under the gauge symmetry $\Psi_{\mu\alpha}\to\Psi_{\mu\alpha}+\del_\mu \chi_\alpha$.  }

Since we will be only interested in zero-frequency correlation functions, it is sufficient for us to consider background fields that are independent of time. We therefore take the metric and gauge connection to have the following form 
\eqn\backgroundfields{\eqalign{&ds^2 = e^{2\phi}(dX^4 + a_i dx^i)^2 + h_{ij}dx^i dx^j~, \cr & A=  A_4 (dX^4 + a_i dx^i) + \CA_i dx^i~. }}
The total space is topologically $\CM_4=\CM_3\times S^1$ and 
$i=1,2, 3$ runs over the coordinates on $\CM_3$. All background fields are taken to be functions of only the $x^i$. $X^4\simeq X^4+\beta$ describes a circle of length $\beta$. To simplify several formulae below, we set $\phi = 0$ (it is straightforward to reintroduce $\phi$).

Introducing a chemical potential $\mu$ in the partition function~\path\ corresponds to the constant background value $A_4 = -i \mu$. Here there is a subtlety that will be important later, so we would like to make a brief digression. Normally, there are two equivalent ways of computing correlation functions in the presence of temperature and chemical potential. One is to introduce the chemical potential as a background field for $A_4$ and the other is to gauge away $A_4$ at the expense of introducing twisted boundary conditions for charged fields. We would like to emphasize that these procedures are not equivalent in the presence of anomalies. Indeed, the required change of variables when going from one description to the other leads to a nontrivial Jacobian. 

The high-temperature limit of the partition function as well as of any other equilibrium observable at zero frequency can be conveniently studied by KK reducing over the $S^1$. Due to our assumption of a gap, we obtain a massive theory on ${\cal M}_3$. Upon integrating out the KK tower, we thus get a {\it local} generating functional for the background fields, ${\cal W}$. This local generating functional for the background fields admits a derivative expansion (which is useful at high temperatures). This is the subject of the next subsection. 

\subsec{High-Temperature Effective Field Theory}

At zeroth order in derivatives we have 
\eqn\zerothorder{{\cal W}^{(0)}= \int_{\CM_3}d^3x \sqrt{h} \, P(A_4,\beta)   }
with an arbitrary function $P$. Actually, in the absence of anomalies, $A_4$ would be a periodic scalar $A_4\simeq A_4+{2\pi \over \beta}$ and so the function $P$ should only depend on $\exp(i \beta A_4)$.

The terms which are first order in derivatives have been classified in~\BanerjeeIZ\
\eqn\firstorder{\CW^{(1)} =  {1 \over r_1}{ik_1 \over 4\pi} \int_{\CM_3} \CA \wedge da + {ik_2r_1 \over 4\pi}\int_{\CM_3} A_4 \CA \wedge d \CA + {i k_3r_1 \over 4\pi}\int_{\CM_3} A_4^2 \CA \wedge da~. }
(In our convention for the metric,~\backgroundfields, the KK photon $a$ is dimensionless. This accounts for the various factors of $r_1$ appearing in~\firstorder.) Note that the coefficients $k_2,k_3$ are associated to field-dependent Chern-Simons terms. These are not standard terms in three-dimensional QFT because they violate gauge invariance. Here we have infinitely many KK fields in three dimensions, so such non-gauge invariant  terms may arise in principle due to the need to regulate the sum over the infinitely many KK fields. 

The sum over the infinitely many KK fields has a preferred regularization. One  requires that the partition function ${\cal W}$ satisfies the four-dimensional anomaly equation (in the equation below we ignore the mixed gauge-gravitational anomaly) 
\eqn\anomalymatchingcubic{A_\mu \to A_\mu  + \partial_\mu \Lambda: \qquad \delta_\Lambda \CW = -i { C \over 24 \pi^2} \int_{\CM_3 \times S^1} \Lambda F \wedge F ~,}
where $C$ is the usual $U(1)^3$ anomaly coefficient, such that $C=1$ for a left-handed fermion of unit charge. Dimensionally reducing the right hand side of~\anomalymatchingcubic\ over the circle, we can match with the gauge variation of~\firstorder\ and find 
\eqn\koneandtwo{k_2=2k_3=- {2\over 3}C~.}
This means that some low momentum correlation functions in thermal field theory are completely determined by the flat-space anomaly coefficients. As has been emphasized in~\BanerjeeIZ,  this provides a microscopic derivation of the chiral magnetic effect of~\SonTF.

Note that such considerations do not fix $k_1$ because it multiplies a term that is invariant under small gauge transformations. This will be the subject of subsection~2.4.

\subsec{Corrections from Anomaly Inflow}

The generating functional constructed in the previous subsection reproduces the four-dimensional anomaly~\anomalymatchingcubic. As is well known~\BardeenPM, this generates correlation functions of the so-called consistent (but non-invariant) currents. Consistent currents are derived from a four-dimensional effective action that satisfies~\anomalymatchingcubic\ by taking a functional derivative with respect to the gauge field. However, as follows directly from~\anomalymatchingcubic, such currents are not themselve gauge invariant.

 If we amend our four-dimensional theory by the five-dimensional term 
\eqn\anomalyinflow{ \delta \CW=i{C\over 24\pi^2}\int_{\CN_5} A^{(5)}\wedge F^{(5)} \wedge F^{(5)}  ~,} 
then the full theory now becomes formally gauge invariant. Here $\CN_5$ is a five manifold such that $\partial \CN_5 = \CM_4$. The superscript $(5)$ signifies that these are some extensions to $\CN_5$ of our background fields. By taking a functional derivative of~\anomalyinflow\ with respect to the gauge field on $\CM_4$, one finds currents which are gauge invariant. One therefore distinguishes the consistent currents from the invariant currents (more generally, the latter are called ``covariant'' currents).
In many explicit calculations in QFT,  the invariant rather than the consistent currents appear naturally. This will also be the case for us.

Since $\CM_4=\CM_3\times S^1$, it is natural to require that $\CN_5=\CN_4\times S^1$ with $\del\CN_4=\CM_3$. (There is no topological obstruction here.) 
The two formalisms that we discussed in subsection 2.1 regarding how to treat the theory in the presence of temperature and chemical potential become equivalent after our theory is amended by~\anomalyinflow.

If we restrict to {\it constant} values for $A_4$,  the dimensional reduction of the $5d$ Chern-Simons term~\anomalyinflow\ gives a total derivative on $\CN_4$. (More precisely, one gets a term that is independent of small deformations of $\CN_4$ or of the gauge connection.) Therefore, it {\it formally} defines a $3d$ local action
\eqn\reductionCS{\eqalign{  \delta {\cal W}\bigr|_{A_4 = {\rm const}} =  -i{C \over 12 \pi}  r_1 \int_{\CM_3} \left( A_4 \CA \wedge d\CA + 2 A_4^2 \CA \wedge da + A_4^3  a \wedge da \right)~.}}

If we now add~\reductionCS\ to~\firstorder\ (and substitute~\koneandtwo) we find the ``covariant'' one-derivative effective action  
\eqn\effectiveaction{\eqalign{  \CW^{(1), cov}\bigr|_{A_4 = {\rm const}}  =& -i{C \over 4 \pi} r_1 \int_{\CM_3} \left( A_4 \CA \wedge d\CA  + A_4^2 \CA \wedge da +{1 \over 3} A_4^3 a \wedge da \right)  \cr&+{1 \over r_1}{ik_1 \over 4 \pi} \int_{\CM_3} \CA \wedge da~.}} 
We emphasize that the $k_1$ term is unaffected by whether we choose to study the ordinary effective action or the ``covariant'' effective action. By construction, functional derivatives of the effective action~\effectiveaction\ correspond to correlation functions of the invariant currents.

In the following subsection we will check that~\effectiveaction\ is indeed obtained from the thermal field theory of a single chiral fermion. In addition, the study of the free fermion will lead to a proof (valid for all Lagrangian theories) that  
\eqn\kgcoeff{k_1 = -{1\over 12}  \Tr(U(1))   ~.}

\subsec{The Thermal Effective Action of a Free Weyl Fermion}
Let us consider  the four-dimensional theory of a massless Weyl fermion $\psi_\alpha$ charged under a $U(1)$ gauge field with charge $e$. We take the space to be topologically ${\cal M}_4={\cal M}_3\times S^1$, with the curvature of ${\cal M}_3$ much smaller than the inverse radius of the $S^1$ (i.e. the KK scale). The fermion is assigned anti-periodic condition along the $S^1$.

The dimensionally-reduced theory on $\CM_3$ is gapped, and the spectrum of the low-energy theory on $\CM_3$ is a tower of fermions, with masses $r_1 m_n=n- e r_1A_4$, where $n\in \Z+1/2$. The tower is coupled to the three-dimensional gauge field ${\cal A}_i$ and also to the graviphoton $a_i$. Under the latter the $n$th particle carries charge $n\in \Z+1/2$.

Recall the following fact about  the 3d theory of a single massive fermion $\psi_\alpha$ with charges $e_x$ under the $U(1)$ gauge fields $A^x$: upon integrating this fermion out, one generates the Chern-Simons terms 
\eqn\geneff{\CW_{eff}=-{i\over 8\pi}{\rm sgn}(m)\int_{\CM_3} \sum e_xe_y A^x\wedge d A^y ~.}

Integrating out the $n$th KK fermion, we thus find (according to~\geneff) the following Chern-Simons terms: 
\eqn\nth{\eqalign{\CW^{(n)}_{eff}  = & -{i\over 8\pi}{\rm sgn}(n- e r_1 A_4)\int_{\CM_3} \left(  e^2 {\cal  A}\wedge d {\cal A}+2e{n \over r_1} {\cal A}\wedge da+{n^2\over r_1^2} a\wedge da \right)~.}}
It is crucial that~\nth\ is (up to the familiar parity anomaly~\refs{\NiemiRQ, \RedlichKN,\RedlichDV}) a correctly quantized, gauge invariant Chern-Simons term.

But now we need to sum over $n\in \Z+1/2$ $$\CW_{fermion}=\sum_{n}\CW_{eff}^{(n)}~.$$
This sum is divergent. We will regulate it using the zeta function. We need the following three sums: 
\eqn\infinitesums{\eqalign{S_1(s,A_4) & =\sum_{n\in\Z+1/2 }{\rm sgn}(n-e r_1A_4) |n-e r_1A_4|^{-s}~,\cr S_2(s,A_4) & =\sum_{n\in \Z+1/2}{\rm sgn}(n-er_1A_4) n |n-er_1A_4|^{-s}~, \cr S_3(s,A_4) & =\sum_{n\in \Z+1/2}{\rm sgn}(n-e r_1A_4) n^2 |n-er_1A_4|^{-s}~,}}
evaluated at $s=0$.
For large enough $s$ all the sums above converge. We take $er_1A_4\in (-\half,\half)$ for simplicity. After some algebra we find 
\eqn\sumresults{\eqalign{ S_1(s=0,A_4) & =2e r_1 A_4~,  \cr
S_2(s=0,A_4) & =e^2r_1^2 A_4^2+1/12~, \cr
S_3(s=0,A_4) & ={2\over 3}e^3 r_1^3 A_4^3~.}}
We thus find the following effective action: 
\eqn\final{ \eqalign{\CW_{fermion} =  -{i\over 4\pi}\int_{\CM_3} & \left(e^3 r_1 A_4  {\cal A}\wedge d{\cal  A}+\left(e^3r_1 A_4^2+{e\over 12 r_1}\right) {\cal A}\wedge da +{e^3 r_1\over 3} A_4^3 a\wedge da  \right)~.}}
This effective action precisely coincides with  \effectiveaction\ (for $C=e^3$) and also agrees with~\kgcoeff.  Zeta-function regularization thus corresponds to  the generating functional for the correlation functions of covariant currents.

Several additional comments:
\item{1.} While the contribution from integrating out each individual field in the KK tower leads to a properly quantized Chern-Simons term, we see that the sum over the KK tower leads to incorrectly quantized (i.e. non-gauge invariant) Chern-Simons terms. 
\item{2.} The term $-{i\over 4\pi}\int_{\CM_3}{1\over 12 r_1}{\cal A}\wedge da$, which is one of the Chern-Simons terms obtained above, was already computed in essentially this way, i.e. by integrating out the KK tower, in~\GolkarKB.  
\item{3.} Let us imagine an arbitrary Lagrangian field theory. (By that we mean that there exists a point in the space of continuous couplings such that the theory becomes free.) If the coefficient of $\int_{\CM_3}{1\over  r_1}{\cal A}\wedge da$ had depended on any continuous couplings, we would have arrived at a contradiction because upon promoting these couplings to background fields we would have violated gauge invariance under small gauge transformations. Since there is no local four-dimensional anomaly to soak up this non-gauge invariance, any dependence on continuous coupling constants is therefore disallowed.  We can thus compute the coefficient of $\int_{\CM_3}{1\over r_1}{\cal A}\wedge da$ at the free field theory point. By 't~Hooft's anomaly matching, the formula~\kgcoeff\ follows for any value of the coupling constants. (One can view this argument as a non-perturbative generalization of~\ColemanZI\ and therefore also of~\GolkarKB.\foot{Note that in~\GolkarKB\ there is a discussion of some situations where there is  dependence on continuous coupling constants. This can only happen if at least one of our assumptions is violated. For example,  the four-dimensional current $j_\mu$ may not correspond to an actual symmetry in the quantum theory (e.g. the axial $U(1)$ of QCD). })
An argument of similar nature has already appeared in a different context in~\ClossetVP.
  
 \item{4.} For non-Lagrangian theories the argument above immediately shows that the coefficient of $\int_{\CM_3}{1\over r_1}{\cal A}\wedge da$ is independent of continuous coupling constants (therefore, the corresponding conductivities are independent of continuous coupling constants).

\item{5.} The two-point correlation functions that stem from~\final\ describe parity-odd terms in various zero-frequency two-point functions involving $j_\mu$ and the energy-momentum tensor in the thermal state of free fermions. Those were studied using conventional thermal field theory one-loop diagrams, for example, in~\LandsteinerSJA\ (and see references therein). Our results for the free fermion completely agree.

\newsec{Supersymmetric Theories}

\subsec{Preliminaries}

An interesting class of observables in supersymmetric theories consists of their partition functions on compact spaces. The simplest case is that of the partition function on $T^d$, which computes (with signs) the number of ground states when the space-like slice is taken to be $T^{d-1}$. This observable can be used as a litmus test for supersymmetry breaking in flat space~\WittenDF.

More recently, the partition functions of supersymmetric theories on spaces which are topologically $S^3\times S^1$ have been studied, starting with~\refs{\RomelsbergerEG,\KinneyEJ}. One can think of these partition functions as counting (with signs) the states of supersymmetric theories whose space-like slice is taken to have the topology of $S^3$.  In the particular case of superconformal field theories, the operator-state correspondence shows that supersymmetric ground states on the round $S^3$ are in one-to-one correspondence with local operators in $\R^4$ sitting in short representations of the superconformal group. 

More generally, supersymmetric theories (possessing an $R$-symmetry) can be studied on spaces of the form ${\cal M}_3\times S^1$ with ${\cal M}_3$ a Seifert manifold~\refs{\KlareGN,\DumitrescuHA,\ClossetRU}. A threefold ${\cal M}_3$ is called Seifert if it can be viewed as a circle fibration over a Riemann surface. The metric on this space has to have a Killing vector with closed orbits.
The partition function on ${\cal M}_3\times S^1$ (with, as required by supersymmetry, periodic boundary conditions for fermions) can be interpreted as 
\eqn\partfun{Z_{{\cal M}_3\times S^1}={\Tr}_{{\cal H}(\CM_3)}\left(e^{-\beta H}(-1)^F\right)~,}
where ${\cal H}(\CM_3)$ is the Hilbert space of the the theory on ${\cal M}_3$ and $H$ is the Hamiltonian which generates motion along the $S^1$.

Under the conditions explained above, spaces of the form ${\cal M}_3\times S^1$ admit two supercharges of opposite charges under the $R$-symmetry. Denoting the corresponding SUSY transformations by $\delta_\zeta$ and $\delta_{\tilde \zeta}$, the superalgebra is 
\eqn\superalgebra{\{\delta_\zeta,\delta_{\tilde \zeta}\}=2i\delta_K~,\qquad K^\mu=\zeta\sigma^\mu\tilde \zeta~.}
One finds that $K^\mu$ is a complex (in fact holomorphic) Killing vector on ${\cal M}_3\times S^1$. One also has  that $\delta_\zeta^2=\delta_{\tilde\zeta}^2=0$. Finally, $\delta_K$ acts as a covariant Lie derivative along~$K^\mu$. 

Since $K$ is a complex Killing vector, it is generally different from $H$.  In this class of SUSY compactifications it is crucial that we have 
\eqn\timeinde{[H,\delta_{\zeta}]=[H,\delta_{\tilde\zeta}]=0~,}
which means that the SUSY parameters are time independent. This also implies that $[H,K]=0$. 
The states that contribute to~\partfun\ are therefore (by the usual argument of decoupling of long representations) annihilated by the operator on the right hand side of~\superalgebra. Thus, only short representations of the superalgebra~\superalgebra\ contribute to~\partfun. One may thus evaluate the partition function at weak (or even vanishing) coupling. In various cases, this has led to remarkable new non-perturbative results about SUSY theories. See for example the work of~\DolanQI.

The dependence on $\beta$ of~\partfun\ is generally nontrivial. This is because the states that contribute to the partition function are annihilated by $K$, but generally not by $H$.

A question that we would like to address here concerns with the $\beta\rightarrow 0$ limit of~\partfun. One of our central findings is the following formula 
\eqn\formula{\beta \to 0: \quad \log Z_{{\cal M}_3\times S^1} =- {\pi^2  \Tr (R)    \over \beta}L_{\CM_3} + {\rm subleading}~,}
where
\eqn\formulaL{\eqalign{ L_{\CM_3}  \equiv {1 \over 48 \pi^2} \int_{\CM_3} d^3x\sqrt{h}  \CR^{(3)} +\cdots ~,}}
and $\cdots$ stand for the additional terms required by supersymmetry. $ \CR^{(3)}$ is the Ricci scalar associated to the metric $h_{ij}$ on ${\cal M}_3$ (we shall use the usual convention, where the curvature of the sphere is positive).

We will now derive~\formula\ and specify the additional terms in~\formulaL. In fact, we will derive a slightly more general form that reduces to~\formula\ in the appropriate circumstances. Consistency checks and applications will be discussed in section 4. 

\subsec{Derivation of the $\beta\to 0$ Limit}

Supersymmetry requires periodic boundary conditions for the fermions as we go around the $S^1$ in ${\cal M}_3\times S^1$. Thus, the main difference with respect to the discussion in section 2 of the thermal partition function is that  now  the low-energy theory in $3d$ has a gapless sector. 

Let us reconsider for a moment the calculation of the thermal partition function of a free fermion~\final\ (we take $e=1$ for simplicity). For $A_4=0$ the fermion has anti-periodic boundary conditions. If we turn on $A_4$, we can shift the masses of the fermions in the KK tower according to $m_n={n\over r_1}-A_4$. If we dial $A_4$ until it eventually reaches $r_1A_4=\pm 1/2$ then we are effectively describing a periodic fermion. (In other words, we are gauging away the flat connection $r_1A_4=\pm 1/2$ at the expense of changing the boundary conditions from periodic to anti-periodic. As explained in subsection 2.3, this is allowed even in the presence of anomalies. Recall that $\CA$ is defined as in~\backgroundfields. Gauging away the half-integer flat connection also leads to a simple redefinition of $\CA$.)  
From~\final\ we see that this shifts the coefficient of $\int_{{\cal M}_3}\CA\wedge da$ from 
${-i\over 48\pi r_1}$ to ${i\over 24\pi r_1}$.\foot{An easier way to arrive at the same conclusion is to consider directly the KK tower associated to a periodic fermion. The coefficient of  $\int_{{\cal M}_3}\CA\wedge da$ then arises from the second sum in~\infinitesums, but now with $n\in\Z$. For $A_4=0$ this sum is just $-1/6$. Compare with~\sumresults.}

Given that there is a massless fermion in the spectrum, does it make sense to discuss this local Chern-Simons term? The answer is certainly positive, for example, because this Chern-Simons term is proportional to $1/r_1$. Indeed, $r_1$ is not a scale in the theory of the massless fermion, so no separated-points correlation function contains $r_1$. Consequently, this Chern-Simons term cannot be contaminated by  nonlocal terms that arise due to the massless fermion. (All the non-local terms are finite in the limit $r_1\to 0$ because they do not contain $r_1$.\foot{In subsection 4.3 we will see a curious but harmless exception to this argument.}) Equivalently, the massless fermion is neutral under the KK photon gauge symmetry and hence cannot contribute non-local terms that depend on $a$.  

In summary, for supersymmetric compactifications on ${\cal M}_3\times S^1$, once we reduce on the $S^1$, there is necessarily the following local term in the generating functional on ${\cal M}_3$: 
\eqn\SUSYlocal{\CW={i \Tr (U(1))\over 24\pi r_1}\int_{\CM_3}\CA\wedge da+\cdots~.}
The gauge field $\CA$ could be a background gauge field for the $R$-symmetry or for some flavor $U(1)$ symmetry; the result~\SUSYlocal\ holds true in both cases (the extent to which this result should be viewed as proven has been discussed in section~2).
In the following subsection we will discuss how to supersymmetrize~\SUSYlocal.  One can also discuss the other terms in~\effectiveaction. Here we just focus on the most singular one,~\SUSYlocal, which is related to the mixed gauge-gravitational anomalies. 

\subsec{Supersymmetrization of~\SUSYlocal\ when $\CA$ is the $R$-Symmetry Gauge Field}

For the case that the gauge field $\CA$ is taken to be the $R$-gauge field, the supersymmetrization of~\SUSYlocal\ proceeds by identifying $\CA$ as a member of the $\CN=2$ $d=3$ supergravity multiplet.
The supergravity multiplet consists of $(h_{ij}, b_{ij}, C_i, \CA^{(R)}_i)$, i.e. the metric, a two-form gauge field $b_{ij}$, a gauge field $C_i$ associated to the central charge, and a gauge field $\CA^{(R)}_i$ that couples to the $R$-current.\foot{This supergravity multiplet is analogous to new-minimal supergravity in four dimensions~\SohniusTP. At the linearized level, the supergravity fields couple to a supercurrent multiplet which contains a conserved $R$-current. For the general formalism see for example~\refs{\KomargodskiRB,\DumitrescuIU}. It has been recently studied at the non-linear level by~\KuzenkoUYA.}

In order to supersymmetrize~\SUSYlocal, we need to identity the KK photon field strength $da$ with one of the fields in the 3d $\CN=2$ supergravity multiplet.  Define the field strength of the gauge field $C_i$ as 
\eqn\fsC{v^i  \equiv -i \epsilon^{ijk} \partial_j C_k ~.}
Then, in the context of dimensional reduction $ \CM_3\times S^1$ over the circle (with some KK photon $a$), it has been shown in~\ClossetRU\ that 
\eqn\dimred{v^i =  -i \epsilon^{ijk} \partial_j a_k~.}

Therefore, in order to supersymmetrize~\SUSYlocal\ we need to find a $d=3$ $\CN=2$ supergravity term that includes $\CA^{(R)}_i v^i$. This is provided simply by the Einstein-Hilbert term\foot{This Lagrangian can be found in several ways, for example, it can be extracted from~\KuzenkoUYA\ or, with a little bit of work, from~\ClossetRU.} 
\eqn\EH{\CL_{EH} = M\left( \half \CR^{(3)} - H^2 + 2 v^i v_i - 2 \CA^{(R)}_i v^i \right)~,}
where $H= -i \epsilon^{ijk} \partial_i b_{jk}$. We have dropped the terms including fermions, since they would not contribute to the partition function when we eventually substitute the appropriate background fields. $M$ is an arbitrary mass scale.  

Since the coefficient of~\SUSYlocal\ is fixed, we find that the scale $M$ is fixed as well. We can summarize  by quoting the contribution of~\EH\ to the $\CM_3\times S^1$ partition function 
\eqn\formula{\beta \to 0: \quad \log Z_{\CM_3\times S^1} = {\pi^2 \kappa L_{\CM_3} \over \beta} + \CO(1)~,}
where
\eqn\formulaL{\eqalign{ L_{\CM_3} & \equiv {1 \over 24 \pi^2} \int_{\CM_3} dx^3\sqrt{h} \left(\half \CR^{(3)} - H^2 + 2 v^i v_i - 2 \CA_i v^i \right)~, \cr \kappa & = - \Tr(R) ~.}}

For any given supersymmetric compactification on a manifold $\CM_3\times S^1$, one now only needs to know the configuration of the background fields appearing in~\formulaL. These have already been fixed in full generality for any complex four-fold of the type $\CM_3\times S^1$ and for the corresponding Seifert manifold~$\CM_3$ in~\refs{\DumitrescuHA,\ClossetRU}. To keep our presentation short, instead of reviewing the results in generality, we will discuss some simple examples in the next section.

\subsec{Supersymmetrization of~\SUSYlocal\ when $\CA$ is a Flavor Gauge Field}

The $3d$ Fayet-Iliopoulos D-term Lagrangian $\sim \xi \int d^4\theta V$ has a simple generalization in the context of curved supersymmetry on ${\cal M}_3$. 

The curved space generalization of the Fayet-Iliopoulos D-term is~\ClossetRU\
\eqn\curvedFI{{\cal L}_{FI}=\xi\left(D-\CA_i v^i-\sigma H\right)~,}
where $\CA$ is the flavor gauge field, $D$ is proportional to the top component of the background vector multiplet in which $\CA$ is embedded, and $\sigma$ is the scalar in the same vector multiplet. $v^i$ and $H$ were defined in the previous subsection.

Using~\SUSYlocal\ and~\dimred\ we infer that in the context of dimensional reduction over the $S^1$ of  ${\cal M}_3\times S^1$ we must have $\xi={\Tr(U(1))\over 12\beta}$. The Fayet-Iliopoulos D-term therefore contributes to the partition function the following singular term as $\beta\to 0$:
\eqn\flavorsingular{\beta \to 0: \quad \log Z_{\CM_3\times S^1} =-{\Tr(U(1))\over 12\beta}\int d^3x \sqrt h \left(D-\CA_i v^i-\sigma H\right)+\cdots~.}
In the following section we will see that the term~\flavorsingular\ can be activated on various supersymmetric configurations, for example, when we turn on a real mass. Therefore, the small circle limit of the partition function contains a term that is fixed by the mixed gauge-gravitational $\Tr(U(1))$ anomaly (for any flavor symmetry).

\newsec{Applications}

\subsec{Hopf Surfaces} 

An interesting example to consider is the partition function over $\CM_4=S^3_b\times S^1$, where $S_b^3$ stands for the squashed three-sphere with parameter $b$. The metric is a product metric with $S^1$ having length $\beta$ and the metric on $S_b^3$ being
\eqn\metricsquashedthree{ds^2_{S_b^3}=r_3^2\left[b^{-2} \cos^2\psi d\phi^2 + b^2 \sin^2\psi d\chi^2 + f(\psi)^2 d\psi^2\right]~,}
with $f(\psi)=\sqrt{b^2\cos^2\psi+b^{-2}\sin^2\psi}$. The range of the angles is $\phi,\chi\in[0,2\pi]$, $\psi\in[0,{\pi\over 2}]$. For $b=1$ $S^3_b$ becomes the usual round sphere.
The total space  $S^3_b\times S^1$ thus has the line element 
\eqn\metricsquashed{ds^2=r_1^2d\theta^2+ds^2_{S_b^3}~,}
with $\theta\simeq\theta+2\pi$. The metric~\metricsquashed\ can be viewed as a Hermitian metric corresponding to a  point on the moduli space of complex structures of  $S^3\times S^1$. (This moduli space is two-complex dimensional.) In terms of the usual\foot{For a review, see~\ClossetVRA.} Kodaira-Spencer coordinates $(p,q)$ we have the relations (we take $b,\beta$ to be real for simplicity)
$p=e^{-\beta b/r_3}$, $q=e^{-\beta b^{-1}/r_3}$. 

The three-dimensional squashed sphere~\metricsquashedthree\ (as well as the four-dimensional space \metricsquashed)  preserves two supercharges. In order to write supersymmetric theories on~\metricsquashedthree\ one needs to activate the background field 
$H=-{i\over r_3 f(\psi)}$ in addition to the metric. The background field $v^i$ vanishes because the four-dimensional metric~\metricsquashed\ is a direct product (see~\dimred).

We are thus ready to compute $L_{S^3_b}$~\formulaL, and we find
\eqn\Lthree{L_{S^3_b}={r_3\over 3}{b+b^{-1}\over 2}~.} 
If there exists a continuous global symmetry we can turn on a real mass and study the effect on the partition function by evaluating the integral in~\flavorsingular. The curved-space version of a real mass corresponds to turning on $\sigma=m$ and~$D=-m H$ (see~\ClossetRU). We then find that for such a configuration of background fields 
\eqn\intflavor{\int d^3x \sqrt h \left(D-\CA_i v^i-\sigma H\right)=-2m\int d^3x \sqrt h H=imr_3^2(2\pi)^2 ~.}
Combining~\Lthree\ and~\intflavor\ we find the asymptotic form of the partition function on the space $S^1\times S^3_b$ 
\eqn\asymform{\beta \to 0: \quad \log Z_{S^3_b\times S^1} =- {\pi^2  r_3(b+b^{-1})   \over 6\beta} \Tr(R)-im{\pi^2 r_3^2\over 3\beta}    \Tr(U(1))+\CO(1)~.}
The equation above is consistent with the fact that the dependence on the $R$-symmetry should be such that a shift of the $R$-current  can be compensated  by some shift of the real mass. The coefficient that relates the shift in the $R$-current and the shift in the real mass can be calculated in complete generality as explained in~\ClossetUDA\ (and see references therein), in agreement with~\asymform.

The supersymmetric index corresponding to the space $S_b^3\times S^1$  has been computed in various theories, and in some cases its $\beta\rightarrow 0$ asymptotics has been explicitly discussed. See for example~\refs{\ImamuraUW,\SpiridonovWW,\AharonyDHA}. Our claim~\asymform\ agrees with these computations.

In the preceding discussions we have not been careful about distinguishing the partition function from the index. On general grounds, we expect that they differ by local counter-terms and by Casimir energy factors (Casimir energy factors are indeed present, see for example~\refs{\KimAVA,\CassaniZWA}). Both of these effects necessarily scale like $\CO(\beta)$, therefore they are inconsequential as far as the singular terms in $\beta$ go. As an example, in the appendix we compute the partition function for the simplest possible case, i.e. a free chiral multiplet on $S^3\times S^1$, and verify~\asymform\ (with $m=0$). Of course, the fact that the chiral multiplet computation gives~\asymform\ is related by supersymmetry to the fact that a free fermion contributes~\mixedCS\ to the generating functional, as we have seen explicitly in section~2.

The computations in section~2 and in the appendix rely on some zeta-function regularization. How do we know they are right? There are several properties that our procedure satisfies. First of all, it produces the same singular terms in the partition function as in the supersymmetric index, which is what one expects as explained above. Second, the result ~\kgcoeff, from which everything we have discussed up to here follows essentially without extra assumptions, has been derived also in different ways and in different situations. See the references in section~2. Third, our regularization procedure for the free fermion field precisely reproduces the first line of~\effectiveaction. The first line of~\effectiveaction\ is fixed by anomalous Ward identities and therefore it must be right. In addition, our method yields the correct result for the singular term of the free energy of two-dimensional conformal field theories, in agreement with modular invariance (we do not present this simple computation here).  Finally, analogous consistency checks on the coefficients of low-energy Chern-Simons terms exist in five dimensions (see the next section for references). Note that the regularization procedure of~\refs{\ClossetSXA,\AsselPAA} disagrees with our results for the singular terms and does not satisfy some of the crucial consistency checks  above. It can however be fixed to agree with our results for the singular terms.\foot{We thank B. Assel, D. Cassani and D. Martelli for several discussions on this.} This and several other related topics will be discussed in~\refs{\AsselInProgress}.

\subsec{Geometric Invariants}

We will now make a few brief comments on a  subject that requires a more thorough treatment.
In~\refs{\ClossetVRA,\ClossetUDA} it was shown that the partition function on some complex  $\CM_4$ does not depend on the Hermitian metric, rather, only on the complex structure of~$\CM_4$. In particular, there is the following family of metrics on $S^3\times S^1$, all of which correspond to the same $p,q$:
\eqn\family{ds^2=r_1^2d\theta^2+r_3^2\left[b^{-2} \cos^2\psi d\phi^2 + b^2 \sin^2\psi d\chi^2 + f(\psi)^2 d\psi^2\right]~, }
but rather than taking $f(\psi)$ to be $f(\psi)=\sqrt{b^2\cos^2\psi+b^{-2}\sin^2\psi}$ as above, we could take {\it any} $f(\psi)$ which approaches $b^{-1}$ at $\psi=\pi/2$ and $b$ at $\psi=0$.
The  background field $H$ is given by $H=-{i\over r_3 f(\psi)}$. The background field $v^i$ vanishes. 
This family of metrics has been first discussed in~\MartelliFU\ (see also~\AldayLBA).

Our claims can therefore be consistent only if the integrated local terms appearing in~\formulaL,\flavorsingular\  do not depend of $f(\psi)$ (except for the values of $f(\psi)$ at the boundaries). Indeed, evaluating the local term~\formulaL\ we find that  
\eqn\evaluation{\eqalign{L_{S^3_b} & \sim  \int_0^{\pi \over 2} d\psi\left( {2\over f(\psi)} \sin(2 \psi) + {\partial_\psi f (\psi) \over f(\psi)^2}\cos(2 \psi) \right) \cr & =  \int_0^{\pi \over 2} d\psi \, \partial_\psi \left(- {1\over f(\psi)} \cos(2\psi)\right) = {1\over f(\pi / 2)} + {1 \over f(0)}=b+b^{-1} ~.}}
Similarly, a real mass term in curved space corresponds to setting $D=-mH$ and one can thus see that in~\flavorsingular\ the dependence on $f(\psi)$ cancels altogether. 

This means that we have found local densities that are invariant under some {\it subset} of the metric deformations of transverse holomorphic foliations. These local densities are therefore somewhat analogous to the familiar topological invariants in even dimensions. A more complete understanding of such invariants is beyond the scope of this paper.\foot{ We hasten to add that while the supersymmetric Einstein-Hilbert density is invariant only under a subset of the metric deformations of transverse holomorphic foliations, similar ideas lead to some other local terms which are apparently completely invariant. We thank C.~Closset and T.~Dumitrescu for illuminating discussions of this. }

\subsec{The Free Vector Field}

So far, when we wrote equations such as~\asymform\ we have implicitly assumed that the massless sector does not produce terms which depend singularly on $r_1$. This assumption can be spoiled if there are infrared divergences, namely, if the massless sector in three dimensions does not have a finite partition function on $\CM_3$. Here we will study the simplest example where this takes place and show that the singular term in $r_1$ is only logarithmic, so the leading singularity in $r_1$ is still governed by~\asymform. We will argue that this is generally the case.

Consider a free vector multiplet in four dimensions. When we take the limit $r_1\rightarrow 0$, the three-dimensional theory that  remains  is the three-dimensional vector multiplet (whose bosonic degrees of freedom are a gauge field and a scalar). The latter theory has an infinite $S^3$ partition function because the scalar in the $3d$ vector multiplet has no curvature coupling. In other words, the Coulomb branch is not lifted. In addition, since the vector field in $d=3$ is dual to a scalar with a  shift symmetry, another divergence arises from the vector field. Therefore, the $\CO(1)$ term in~\asymform\ actually diverges! If we keep $r_1$ finite but small, then the scalar in the three-dimensional vector multiplet has a finite radius (due to large four-dimensional gauge transformations) and the gauge coupling of the three-dimensional vector field is finite. The  divergence as $r_1\rightarrow 0$ is due to the fact that the moduli space becomes non-compact, and hence it appears logarithmically in $\log Z$. The $1/r_1$ term is therefore still present as claimed in~\asymform\ but the $\CO(1)$ term is actually a logarithm of $r_1$. Indeed, the superconformal index of a free four-dimensional vector multiplet is given by
\eqn\indexvec{\CI_{\rm vector}(p,q) = \prod_{l=1}^{\infty}(1-p^l)(1-q^l)~,}
and expanding this expression in the limit  $r_1\rightarrow 0$ we find
\eqn\limitfreevec{r_1\to 0: \quad \log \CI_{\rm vector}(p,q) = -{\pi \over 6}{b + b^{-1} \over 2} {r_3 \over r_1} - \log\left({r_1 \over r_3}\right) + \CO(1)~.}
The leading term nicely agrees with~\asymform. Concerning the logarithmic piece, a single scalar in three dimensions contributes such a logarithmic term with coefficient $-\half$, and since here we have  two such scalars (one comes from the holonomy in four dimensions and the other from dualizing the three-dimensional gauge field), the coefficient of the logarithm is precisely as expected.

More generally, for gauge theories with a Coulomb branch which is not lifted by 3d classical or quantum corrections, we expect the Coulomb branch to be always compact if $r_1$ is kept finite, and thus the partition function only diverges as a power law related to the number of moduli. This only affects $\log Z$ logarithmically in $r_1$.

\subsec{${\cal N}=1$ Superconformal Field Theories }
 It is interesting to specialize~\asymform\ to the case of $\CN=1$  superconformal field theories. We can choose the $R$-symmetry to be the superconformal one and turn off the mass terms in~\asymform. Using the relation to the trace anomalies~\AnselmiAM\  
\eqn\relationanomalies{\Tr(U(1)_R)=16(a-c)~,}
we can rewrite the asymptotic form of the partition function as 
\eqn\scft{\beta \to 0: \quad \log Z_{S^3_b\times S^1} =- {8 \pi^2  r_3(b+b^{-1})   \over 3\beta} (a-c)+\cdots~.}
Radial quantization allows us to reinterpret this partition sum as counting local operators in $\R^4$ that sit in short representations of the superconformal group. When these short representations are counted with signs, one finds a quantity that does not depend on continuous coupling constants~\KinneyEJ. 
The parameter $\beta$ then couples to the Hamiltonian $H=\Delta+\half R$ (where $\Delta$ is the scaling dimension) and $b$ couples to one of the Cartan generators of the angular momentum $SO(4)$. One can view~\scft\ as an exact result for the asymptotics of short representations of $\CN=1$ SCFTs.

Formally, in the limit $\beta \to 0$, the partition function reduces to the unrefined sum over short representations $\Tr((-1)^F)$ (i.e.~the total Witten index). One learns from~\scft\ that

\item{1.)} If $a-c<0$ then fermionic and bosonic operators do not cancel against each other asymptotically, and the total Witten index in the space of local operators therefore diverges. The (absolute value of the) spectral density is asymptotically growing exponentially. 
\item{2.)} If $a=c$ then there is a delicate albeit imperfect cancelation between bosonic and fermionic short representations. The spectral density does not grow exponentially asymptotically. 
\item{3.)} If $a-c>0$ then there is a perfect cancelation between fermionic and  bosonic short representations.  The spectral density is asymptotically oscillatory and the bigger $a-c$ is, the more frequent the oscillations are. 

\medskip

 When one studies the BPS conditions on local operators in $\R^4$, there is no a priori relation between short representations which are fermionic and short representations which are bosonic. Since there are generically infinitely many short representations which are bosonic and infinitely many short representations which are fermionic, one would expect that the total Witten index is generically infinite. 
Thus, the scenario that $a - c > 0$ might seem unlikely or non-generic from this point of view. This could explain why it is much more difficult to construct examples with $a-c>0$, although clearly not impossible (e.g.~the free vector field of the previous subsection).

Note that when $a=c$ we find a vanishing coefficient for the three-dimensional Einstein-Hilbert term in the effective action on $S^3$. Examples of SCFTs with $a=c$ include the theories with $\CN=4$ supersymmetry.  This perhaps suggests that it could be impossible to complete the Einstein-Hilbert term to an action preserving the extended (off-shell) supersymmetry.

The connection between the sign of $a-c$ and the asymptotic structure of short representations is reminiscent of~\KutasovSV. Let us also point out that a different relation between the superconformal index and the Weyl anomalies $a,c$ has been recently discussed in the context of $\CN=2$ superconformal theories in \BuicanQLA. 

Interestingly, in the context of CFTs with a  holographic dual, $a-c$ plays a crucial role as an order parameter for new primary operators of high spin~\CamanhoAPA. It would be nice to understand if there is a relation to our results.

 \newsec{Six Dimensions}

In this section we will outline the generalization to field theories in $6d$. This will lead us to propose a relation between anomalies and the asymptotic behavior of the supersymmetric index in $(1,0)$ theories. 

Let us start by discussing the thermal partition function for a non-supersymmetric $6d$ theory. This is given by the Euclidean path integral over a manifold of the type $\CM_6 = \CM_5 \times S^1$, with anti-periodic conditions for the fermions along the circle. When the theory has a certain global symmetry group $G$, we can couple it to background gauge fields $A_\mu$ and consider the partition function as a functional of the background metric $g_{\mu\nu}$ and of $A_\mu$. The theory at finite temperature is generically gapped. In the limit of small $\beta = 2 \pi r_1$ we can write a local effective action for the (dimensionally reduced) sources on $\CM_5$.

Importantly, the $5d$ effective action must match the anomalous variation of the generating functional of the $6d$ theory. Like in $4d$, there can be anomalies in background $G$ gauge transformations and mixed $G$-gravitational anomalies. Moreover, in $6d$ there can be purely gravitational anomalies. The eight-form anomaly polynomial is \refs{\AlvarezGaumeIG,\AlvarezGaumeDR}
\eqn\anopolsix{I_8 = {1 \over (4 \pi)^3} \left[ {A \over 720} \tr R^4 + {B \over 576} (\tr R^2)^2 - {C\over 12} \tr R^2 \Tr F^2 + {D \over 3} \Tr F^4 \right]~,}
where $\tr$ denotes the trace over Lorentz indices and $\Tr$ over the fundamental representation of $G$. The normalization is such that $A=B=C=D=1$ for a left-handed fermion in the fundamental representation of $G$. The anomalous variation of the generating functional is fixed by the polynomial via the usual descent equations $I_8 = d I_7$, $\delta I_7 = d I_6$ and $\delta \CW = - i \int_{\CM_6} I_6$.

The terms in the $5d$ effective action which are fixed by the anomalies will take the form of $5d$ Chern-Simons terms. Some of them are higher-dimensional analogs of the $k_{2,3}$-terms in \firstorder, namely they are not invariant under small diffeomorphisms/$G$-gauge transformations, and their variation reproduces the anomaly. By dimensional analysis, such terms do not lead to singular terms in the partition function as  $r_1\to 0$. We therefore do not discuss such terms any further. The other, more interesting, terms are the analogs of the $k_1$-term in \firstorder. These Chern-Simons terms are invariant under small diffeomorphisms/$G$ gauge transformations (but not correctly quantized in general) and come with negative powers of $r_1$.

At order $r_1^{-3}$ there is $\sim \int a\wedge da \wedge da$, where $a$ is the KK photon. This is gauge invariant under small KK gauge symmetry transformations. As in our discussion in section~2, the coefficient of this term is expected to be fixed by the purely gravitational anomalies $A$ and $B$. At order $r_1^{-1}$ there are two possible Chern-Simons terms that are invariant under small gauge transformations, $\sim\int a \wedge \Tr(\CR^{(5)}\wedge \CR^{(5)})$ and $\sim\int \Tr (\CA\wedge d\CA -i {2 \over 3} \CA\wedge \CA\wedge \CA) \wedge da$, where $\CR^{(5)}$ is the curvature two-form on $\CM_5$ and $\CA$ is the dimensional reduction of the $G$ gauge field. The former Chern-Simons term is expected to be fixed by the gravitational anomalies $A$ and $B$, while the latter is expected to be fixed by the mixed anomaly $C$.\foot{Like in $4d$, such Chern-Simons terms arising from dimensional reduction are associated to non-dissipative transport coefficients in hydrodynamics. The relations between these transport coefficients and anomalies for $6d$ theories of free Weyl fermions appeared in \LoganayagamPZ~and an argument for their general validity was given in \JensenRGA.}  
(Note that in $5d$, in addition to massive fermions, also massive tensor (two-form) fields generate such Chern-Simons term when they are integrated out~\refs{\BonettiFN,\BonettiELA}. This is necessary for the  relations below between Chern-Simons terms and anomalies to make sense.)  In the small radius limit the thermal partition function will contain terms at order $r_1^{-5}$ and additional terms at orders $r_1^{-3}$ and $r_1^{-1}$ which are not fixed by the anomalies. However, in the presence of supersymmetry, $r_1^{-5}$ would be absent while the other singular terms will be related by supersymmetry to the Chern-Simons terms discussed above. Therefore, the singular terms in $r_1$ will be all fixed by the anomalies $A,B,C$.

We will only consider  $6d$ $(1,0)$ supersymmetric theories in the background $\CM_6 = S^5\times S^1$. Various squashings of the five-sphere preserving an $SU(3)\times U(1)$ isometry and two or more supercharges have been studied in \refs{\ImamuraXG,\ImamuraXNA,\LockhartVP,\AldayRXA,\AldayBTA}. The partition function on this background computes the index \refs{\ImamuraXNA, \LockhartVP,\SpiridonovDE,\BhattacharyaZY}
\eqn\indexsix{Z_{S^5 \times S^1} = {\Tr} _{\CH(S^5)} \left((-1)^F p^{C_1} q^{C_2} t^{C_3} \right).}
Here $C_{1,2} = {1 \over 3}(j_1 \pm {j_3 \over 2}) + {j_2 \over 2} + {R \over 2}$ and $C_3 ={1 \over 3}(j_1 +j_3) +{R \over 2} $, where $j_1$ is the generator of $U(1)$, $j_{2,3}$ are Cartan generators of $SU(3)$ in a suitable basis and $R$ is the Cartan of the $SU(2)_R$ $R$-symmetry (we are using the notation of \ImamuraXNA). We parametrize the fugacities $(p,q,t)$ as
\eqn\pqt{p = e^{- 2\pi {r_1 \over r_5} \omega_1}\,,\quad q= e^{- 2\pi  {r_1\over r_5} \omega_2}\, , \quad t = e^{- 2 \pi  {r_1\over r_5} \omega_3}~,}
where $\omega_{1,2,3} \in \C$ with $\Re(\omega_{1,2,3}) > 0$ and $r_1$ ($r_5$) is the radius of the $S^1$ ($S^5$). In the round case $\omega_1 = \omega_2 = \omega_3 = 1$. We expect $(\omega_1,\omega_2, \omega_3)$ to be expressed in terms of the squashing parameters of the five-sphere. In some examples, the precise relation can be found in~\refs{\ImamuraXNA}. The index \indexsix~can be further generalized to include fugacities for the possible flavor symmetries of the theory.

In the limit $r_1 \to 0$ with $\omega_{1,2,3}$ fixed, the leading behavior of the index \indexsix~is 
\eqn\limitindsix{\eqalign{r_1 \to 0: & \quad  \log Z_{S^5 \times S^1} = - { \pi \over \omega_1 \omega_2 \omega_3 } \left( {\kappa_1 \over 360 }{r_5^3 \over r_1^3} \right. \cr & \left.  +  {(\omega_1^2 + \omega_2^2 + \omega_3^2) \kappa_2 + 3 (\omega_1 \omega_2 + \omega_2 \omega_3 + \omega_3 \omega_1)\kappa_3 \over 72 }{r_5 \over r_1} + \CO(1) \right),} }
where $\kappa_{1,2,3}$ are theory-dependent coefficients, and the normalization is such that $\kappa_1 = \kappa_2 = \kappa_3 = 1$ for a free vector multiplet. 

We propose that 

\item{1.)} The terms at order $r_1^{-3}$ in the five-dimensional effective action are related by supersymmetry to the Chern-Simons term $a\wedge da \wedge da$. The evaluation of the corresponding supersymmetric action on the (squashed) five-sphere gives the $r_1^{-3}$ divergent term in~\limitindsix. This implies that $\kappa_1$ is fixed by the anomaly coefficients $A$ and $B$.
\item{2.)} All the terms at order $r_1^{-1}$ in the five dimensional effective action are related by supersymmetry either to  $ a \wedge \tr(\CR^{(5)}\wedge \CR^{(5)})$ or to $ \Tr (\CA\wedge d\CA -i {2 \over 3} \CA\wedge \CA\wedge \CA) \wedge da$. If we fix $\CA = \CA^{(R)}$ to be the dimensional reduction of the $SU(2)_R$ gauge field, the evaluation of the two corresponding supersymmetric actions gives the two independent $r_1^{-1}$ divergent terms in \limitindsix. This implies that a combination of $\kappa_2,\kappa_3$ is fixed by $A$ and $B$, while an independent combination is fixed by the coefficient $C$ of the $SU(2)_R$ $R$-symmetry. (If the index is generalized to include fugacities for flavor symmetries, we expect additional divergences at order $r_1^{-1}$, whose coefficient will be determined by the $C$ coefficient of the corresponding symmetry. Here we consider only the simplest case with no fugacities for flavor symmetries.)

\medskip

Concretely, our proposal is
\eqn\conjecture{\kappa_1 = -{A + 5 B \over 6}\,,\quad \kappa_2 - {3\over 2} \kappa_3 = {A + 2 B  \over 6}\,,\quad \kappa_3 = -C ~.}
The coefficients in~\conjecture\ agree with the example of a free theory of $n_H$ hypermultiplets, $n_V$ vector multiplets, and $n_T$ tensor multiplets. In this case we have
\eqn\asymptfree{\kappa_1 = - n_H + n_V + n_T \, , \quad \kappa_2 = {n_H + 2 n_V + 2 n_T \over 2}\,,\quad \kappa_3 = n_V - n_T ~,}
while the anomaly coefficients are given by
\eqn\anomfree{A = n_H - n_V + 29 n_T\,,\quad B = n_H - n_V - 7 n_T\,,\quad C= -n_V + n_T~.}
The facts that in free theories $\kappa_1$ only depends on $A$ and $B$, and there is a combination of $\kappa_{2,3}$ which only depends on $C$, are non-trivial checks of our proposal. Since the anomaly coefficients, the various Chern-Simons terms,  and the index are invariant under RG flows, the conjecture automatically holds true for all theories that are connected to free ones by RG flows. 

As a simple test of our conjecture for a theory not connected to any free theory, we consider non-abelian $(2,0)$ theories. The analogy with the maximally supersymmetric case in $d=4$ (see the comment in subsection 4.4), and consistency with dimensional reduction from $6d/5d$ to $4d/3d$, suggest that the leading divergence $r_1^{-3}$ must be always absent in this case. This is in agreement with our proposal~\conjecture, because $A = -5B$ for this class of theories~\refs{\HarveyBX,\IntriligatorEQ}. Various proposals for the computation of the $S^5 \times S^1$ index in these theories have been put forward in the literature, see for instance \refs{\LockhartVP,\KimAVA,\KimQF,\KimNVA}. In \refs{\KimAVA,\KimQF,\KimNVA} the small $r_1$ limit is also discussed, and the leading $r_1^{-3}$ divergence is indeed found to be absent, in agreement with our conjecture. 

An ab initio proof of the relations \conjecture~would require supersymmetrizing the Chern-Simons terms in the appropriate off-shell formulation of $\CN= 1$, $5d$ supergravity, and evaluating the corresponding actions on the supersymmetric $S^1 \times S^5$ background.  We leave this task for future work. Once these off-shell supersymmetric actions are available, the relation between the anomalies of the six-dimensional theory and the $r_1 \to 0$ asymptotics of the partition functions could be readily generalized to all the possible supersymmetric backgrounds of the type $\CM_5 \times S^1$. Analogously to the four-dimensional case, for $6d$ superconformal theories one could rewrite some of the anomalies above in terms of the conformal anomalies (see~\MaxfieldAW\ for a recent discussion in the context of $(2,0)$ theories). 

\vfill\eject
\centerline{\bf Acknowledgments}

We are very grateful to O.~Aharony, A.~A.~Ardehali, B.~Assel, M.~Buican,  D.~Cassani, C.~Closset, T.~Dumitrescu,  D.~Gaiotto, C.~Hoyos, K.~Intriligator, D.~Kutasov, J.~T.~Liu, D.~Martelli, F.~Porri, L.~Rastelli, A.~Schwimmer, I.~Shamir, P.~Szepietowski, and A.~Yarom  for useful discussions. LDP and ZK are grateful to the KITP for its warm hospitality during the initial stages of this project, which was supported in part by the National Science Foundation under Grant No. NSF PHY11-25915. ZK  also thanks the Perimeter Institute for its very kind hospitality during the course of this project. Research at the Perimeter Institute is supported in part by the Government of Canada through NSERC and by the Province of Ontario through MRI. LDP and ZK are supported by the ERC STG grant number 335182, by the Israel Science Foundation under grant number 884/11, by the United States-Israel Binational Science Foundation (BSF) under grant number 2010/629, and by the I-CORE Program of the Planning and Budgeting Committee and by the Israel Science Foundation under grant number 1937/12. Any opinions, findings, and conclusions or recommendations expressed in this material are those of the authors and do not necessarily reflect the views of the funding agencies.

\appendix{A}{The Partition Function of a Free Chiral Multiplet on $S^3\times S^1$}

We consider the simple example of a free chiral superfield on $S^3 \times S^1$. In this case the partition function can be computed explicitly (without using localization). The supersymmetric Lagrangian for a free chiral multiplet of $R$-charge $R$ on this background can be found for example in~\FestucciaWS\
\eqn\freechir{\eqalign{   \CL & =  |\partial_i \phi|^2  -i \bar{\psi} \gamma^i \nabla_i  \psi - \bar{\psi} \partial_4 \psi + {2R -1 \over2 r_3} \bar{\psi}\psi\cr  & + |\partial_4 \phi|^2 + {R-1 \over r_3}(\phi^* \partial_4 \phi - \phi \partial_4 \phi^*) -{R(R-2)\over r_3^2}\phi^*\phi ~,}}
where $i=1,2,3$ labels coordinates on the $S^3$ and $4$ is the coordinate along the circle. Expanding $\psi$ and $\phi$ in Fourier modes on the $S^1$, we get
\eqn\Fourier{\eqalign{\CL & = \sum_{n \in \Z} \left( |\partial_i \phi^{(n)}|^2 -i \bar{\psi}^{(n)} \gamma^i \nabla_i \psi^{(n)} + m_{B,n}^2 |\phi^{(n)}|^2 -i m_{F, n} \bar{\psi}^{(n)} \psi^{(n)}          \right)~. }}
In units of the radius $r_3$ of the three-sphere, the masses of the $n$-th bosonic and fermionic mode are
\eqn\mn{m^2_{B,n} = n^2{ r_3^2 \over r_1^2} + 2 i n(R-1) {r_3 \over r_1 } - R(R-2) \,,\quad m_{F,n} = n { r_3 \over r_1} + i (R - 1/2)~.}
For every mode $(\phi^{(n)}, \psi^{(n)})$, we diagonalize the Laplacian and the Dirac operator on $S^3$ and write the partition function as an infinite sum
\eqn\partfunct{\log Z^{(n)} = - \sum_{l=1}^{\infty}\left(l^2 \log\left[l^2 -1 + m_{B,n}^2\right] - l(l+1)\log\left[\left(l+1/2\right)^2 +  m_{F,n}^2\right] \right).}
This sum is divergent. Following \KlebanovGS~we regularize the sum by shifting $m^2_{B,n} \to M^2_{B,n} = m^2_{B,n} + \mu^2$ and $m^2_{F,n} \to M^2_{F,n} = m^2_{F,n} + \mu^2$ and taking a derivative with respect to $\mu^2$. The derivative of the series can be summed using zeta function regularization
\eqn\regsum{\eqalign{    {\partial \log Z^{(n)} \over \partial \mu^2}& = {\pi \over 2}\sqrt{ M_{B,n}^2 - 1}\, \coth\left(\pi \sqrt{ M_{B,n}^2 - 1} \right) \cr &  - {\pi \over 8} {1 \over \sqrt{M_{F,n}^2}} \left( 4  M_{F,n}^2 + 1\right) \, \tanh\left( \pi  \sqrt{M_{F,n}^2}\right).   }}
Since we are interested in the limit $r_1 / r_3 \to 0$, we expand this expression at large $M_{B,n}$, $ M_{F,n}$
\eqn\taylexp{{\partial \log  Z^{(n)} \over \partial \mu^2} \sim {\pi \over 4}\left(  { 2  M_{B,n}^2 - 1 \over \sqrt{M_{B,n}^2}} - \half {4 M_{F,n}^2 + 1 \over \sqrt{M_{F,n}^2}}\right) + \dots~, }
where the dots denote subleading terms. We can then integrate back  $\mu^2$ and take $\mu^2 = 0$ (we set the $r_1 / r_3$-independent integration constant to zero), with the following result
\eqn\int{ \log Z^{(n)} \sim {\pi \over 6}\left( \sqrt{m_{B,n}^2} (2  m_{B,n}^2 -3) - \half \sqrt{m_{F,n}^2}(4 m_{F,n}^2 + 3)\right) + \dots~.}
Plugging \mn~we see that, as expected from supersymmetry, the volume term ${r_3^3 / r_1^3}$ cancels. The leading terms are
\eqn\expan{\log Z^{(n)} \sim {- i {\pi n |n| \over 2  } {r_3^2 \over r_1 ^2 } + \pi |n| (R -1) {r_3 \over r_1} + \dots~. }}
Finally, we regularize the sum over the KK modes labeled by $n \in \Z$. Using that $\sum_{n \in \Z} n |n| = 0$ and $\sum_{n\in\Z} |n| = -{1 \over 6}$ we obtain the following result for the leading behavior of the partition function 
\eqn\result{\log Z_{S^3 \times S^1} = \sum_{n\in\Z} \log Z^{(n)}  = {\pi \over 6} {r_3 \over r_1}(1 - R) + \dots~. }
This result agrees perfectly with the prediction \asymform.  As we stressed in the main body of the text, such singular terms are scheme independent because no local 4d term can contribute singular terms in $r_1$.

\listrefs

\bye